\documentclass[12pt,aps,preprint,nofootinbib,superscriptaddress,nobalancelastpage]{revtex4}

\usepackage{mathrsfs}
\usepackage{amsmath,amsthm,amssymb,slashed}
\usepackage{axodraw4j}
\usepackage{color}
\usepackage{graphicx}
\usepackage{verbatim}
\usepackage{hyperref}
\hypersetup{colorlinks,bookmarksopen,bookmarksnumbered,citecolor=blue,
linkcolor=blue,pdfstartview=FitH,urlcolor=blue}

\newcommand{\be}{\begin{equation}}
\newcommand{\ee}{\end{equation}}
\newcommand{\bea}{\begin{eqnarray}}
\newcommand{\eea}{\end{eqnarray}}

\newcommand{\Ref}{Ref.}

\newcommand{\eq}{Eq.}

\newcommand{\hc}{\mathrm{H.c.}}

\begin{document}
\preprint{FTUAM-15-22, IFT-UAM/CSIC-15-081, SISSA 32/2015 FISI, LPT-Orsay-15-51}

\title{Loop level constraints on Seesaw neutrino mixing}

\author{Enrique Fernandez-Martinez}
\email{enrique.fernandez-martinez@uam.es}
\affiliation{Departamento de F\'isica Te\'orica, Universidad Aut\'onoma de Madrid, Cantoblanco E-28049 Madrid, Spain}
\affiliation{Instituto de F\'isica Te\'orica UAM/CSIC,
 Calle Nicol\'as Cabrera 13-15, Cantoblanco E-28049 Madrid, Spain}
\author{Josu Hernandez-Garcia}
\email{josu.hernandez@uam.es}
\affiliation{Departamento de F\'isica Te\'orica, Universidad Aut\'onoma de Madrid, Cantoblanco E-28049 Madrid, Spain}
\affiliation{Instituto de F\'isica Te\'orica UAM/CSIC,
 Calle Nicol\'as Cabrera 13-15, Cantoblanco E-28049 Madrid, Spain}
\author{Jacobo Lopez-Pavon}
\email{jacobo.lopezpavon@sissa.it}
\affiliation{SISSA, via Bonomea 265, 34136 Trieste, Italy.}
\affiliation{INFN, sezione di Trieste, 34136 Trieste, Italy.}
\author{Michele Lucente}
\email{michele.lucente@th.u-psud.fr}
\affiliation{Laboratoire de Physique Th\'eorique, CNRS -- UMR 8627, 
Universit\'e de Paris-Sud, F-91405 Orsay Cedex, France.}
\affiliation{SISSA, via Bonomea 265, 34136 Trieste, Italy.}

\begin{abstract}
We perform a detailed study of the importance of loop corrections when deriving bounds on heavy-active neutrino mixing in the context of general Seesaw mechanisms with extra heavy right-handed neutrinos. We find that, for low-scale Seesaws with an approximate $B-L$ symmetry characterized by electroweak scale Majorana masses and large Yukawas, loop corrections could indeed become relevant in a small part of the parameter space. Previous results in the literature showed that a partial cancellation between these important loop corrections and the tree level contributions could relax some constraints and lead to qualitatively different results upon their inclusion. 
However, we find that this cancellation can only take place in presence of large violations of the $B-L$ symmetry, that lead to 
unacceptably large contributions to the light neutrino masses at loop level. Thus, when we restrict our analysis of the key observables 
to an approximate $B-L$ symmetry so as to recover the correct values for neutrino masses, we always find loop corrections to be negligible 
in the regions of the parameter space preferred by data. 
\end{abstract}

\maketitle

\section{Introduction}
\label{sec:intro}

The origin of the observed pattern of neutrino masses and mixings in neutrino oscillation experiments (see e.g. Ref.~\cite{Gonzalez-Garcia:2014bfa} for a recent summary) comprises one of the few experimental evidences for physics beyond the Standard Model (SM) of particle physics. The simplest and most popular extension to account for these experimental observations consists in the addition of right-handed neutrinos to the SM particle content. Given their singlet nature, a Majorana mass term for the right-handed neutrinos is directly allowed in the Lagrangian, thus inducing a new mass scale -the only one unrelated to electroweak (EW) symmetry breaking- to be determined by data. Depending on the size of this scale its phenomenological consequences are very different. 

One of the most appealing choices is that this new Majorana scale is high, leading to the well-known Seesaw mechanism~\cite{Minkowski:1977sc,Mohapatra:1979ia,Yanagida:1979as,GellMann:1980vs} and providing a rationale for the extreme smallness of neutrino masses when compared to the rest of the SM fermions and the EW scale. Values for the neutrino Yukawa couplings ranging between that of the electron and that of the top quark would lead to Majorana masses between the EW and the grand unification scale. Unfortunately, even for the lightest mass choice, any phenomenological consequence beyond neutrino masses tends to be hopelessly suppressed if the extra degrees of freedom only couple to the SM through their Yukawa interactions.

However, the smallness of neutrino masses could derive from symmetry arguments~\cite{Mohapatra:1986bd,Bernabeu:1987gr,Branco:1988ex,Buchmuller:1990du} rather than a hierarchy of scales. Indeed, the Weinberg operator~\cite{Weinberg:1979sa} leading to neutrino masses in Seesaw mechanisms is protected by the $B-L$ symmetry, conserved in the SM and violated in two units by the Weinberg operator. Thus, if the pattern of the Yukawa couplings and Majorana masses in a Seesaw realization is such that it conserves $B-L$, the Weinberg operator will never be generated and the SM neutrinos will remain massless, even for $Y_\nu \sim 1$ and Majorana masses of the order of the EW scale. Small violations of $B-L$ in this pattern would thus induce the small neutrino masses observed in oscillation experiments. In this class of models fall the popular inverse~\cite{Mohapatra:1986bd,Bernabeu:1987gr} or linear~\cite{Malinsky:2005bi} Seesaw mechanisms which, contrary to the canonical type-I Seesaw, would lead to an extremely rich phenomenology through the large mixing allowed between the new extra sterile neutrinos and their SM siblings implying observable contributions in lepton flavour violating (LFV) processes, universality violation and signals in electroweak precision observables. It is then of interest to fit all these available data to determine the allowed values of the mixing of the heavy neutrinos with the SM charged leptons, examples of these constraints can be found in Refs.~\cite{Langacker:1988ur,Bilenky:1992wv,Nardi:1994iv,Tommasini:1995ii,Bergmann:1998rg,Loinaz:2002ep,Loinaz:2003gc,Loinaz:2004qc,Antusch:2006vwa,Antusch:2008tz,Alonso:2012ji,Abada:2012mc,Akhmedov:2013hec,Basso:2013jka,Abada:2013aba,Antusch:2014woa,Antusch:2015mia,Drewes:2015iva}.

When deriving such constraints on heavy-active neutrino mixing, it was recently pointed out in~\cite{Akhmedov:2013hec} that loop corrections involving the extra heavy neutrinos played an important role, obtaining qualitatively different results to those derived by staying at leading order. In particular, it was shown that corrections to the $T$ parameter~\cite{Peskin:1990zt,Peskin:1991sw} could be sizable and that these, in turn, would affect the determination of $G_F$ through $\mu$ decay competing with the tree level effects. Since the value of $G_F$ from $\mu$ decay is generally in good agreement with the measured value of $M_W$ and other determinations of $\sin \theta_W$, in~\cite{Akhmedov:2013hec} it was found that the constraints stemming from these datasets could be weakened at loop level through partial cancellations between the tree level corrections and the $T$ parameter contribution. Furthermore, the invisible width of the $Z$, which is in slight tension with the SM prediction, is modified at tree level through the presence of extra heavy neutrinos, while the oblique corrections computed in~\cite{Akhmedov:2013hec} were found to be subleading. Thus, by accounting for these loop corrections, good fits with relatively large heavy-active mixing were found in~\cite{Akhmedov:2013hec}, since it is possible to alleviate the tension in the invisible width of the $Z$ without seriously affecting the determination of $G_F$ in $\mu$ decay through the partial cancellation of the tree and loop level contributions. 

However, when Ref.~\cite{Basso:2013jka} also investigated the relevance of the $T$ parameter the same cancellation was not reproduced and 
in~\cite{Antusch:2014woa} it was argued that loop contributions should always be negligible, since the heavy-active mixing that controls the strength 
of the couplings of the new degrees of freedom is bounded to be small ($\theta^2 \lesssim 10^{-2}$). Therefore, new tree-level bounds were derived instead 
through more updated fits to available data. While this argument is generally true, models based on an approximate $B-L$ symmetry are characterized by large 
Yukawas and EW-scale Majorana masses, thus, even if loop corrections through weak interactions further suppressed by $\theta^2$ are indeed negligible for all 
practical purposes, when the loop corrections are mediated by heavy neutrinos and/or the Higgs field or its Goldstones, the coupling involved in the vertex is 
no other than the large Yukawa coupling, so that loop corrections can indeed become relevant, as stated in~\cite{Akhmedov:2013hec}. However, not only the oblique 
corrections computed in~\cite{Akhmedov:2013hec} fall in this category, since the effect of the large Yukawa interactions does not vanish in the limit of massless 
neutrinos and charged leptons. Indeed, some vertex and box corrections involving large Yukawas are found not to vanish in the massless limit for light leptons 
(see e.g.~\cite{Kniehl:1996bd}).

In this work we clarify the importance of loop contributions to the determination of the heavy-active neutrino mixing including all loop corrections 
mediated by the potentially large Yukawa interactions. We find that, as discussed by~\cite{Akhmedov:2013hec}, the most relevant of these corrections are 
indeed the ones encoded through the oblique parameters but, in order to make them competitive with the tree-level contributions, EW scale Majorana masses and 
Yukawas on the very border of perturbativity are simultaneously required. Furthermore, we find that, as long as $B-L$ is conserved, the $T$ parameter is always
 positive, so that the partial cancellation discussed in~\cite{Akhmedov:2013hec} cannot take place in such a setup. Large violations of $B-L$ are thus required 
to achieve the negative and sizable values of $T$ capable of reproducing the effect. But these large violations of $B-L$ would render the Weinberg operator 
unprotected and, in presence of the EW-scale Majorana masses and large Yukawas required for $T$, radiative corrections lead to unacceptably large contributions 
to the light neutrino masses, even if these are fixed to their correct value at tree level by means of the Casas-Ibarra parametrization. Thus, when we impose an 
approximate $B-L$ symmetry with only 3 extra heavy right-handed neutrinos, we always find that loop corrections are irrelevant when deriving bounds on the 
heavy-active neutrino mixings. 

This paper is organized as follows: In Section~\ref{sec:param} we detail the parametrization employed for our study. In Section~\ref{sec:obs} we list the observables we analyze in our global fits. In Section~\ref{sec:res} we present our findings and discuss the importance of loop effects in the global fits as well as the necessity of large violations of $B-L$ in order to obtain partial cancellations between the tree and loop level contributions. Finally, in Section~\ref{sec:res} we summarize our results and present our conclusions.
 
\section{Parametrization}
\label{sec:param}

In this work we explore the constraints that can be derived through various EW observables on the extra neutrino mass eigenstates mixing with charged leptons in a Seesaw scenario:
\begin{eqnarray}\label{eq:The3FormsOfNuMassOp}
\mathscr{L} &=& \mathscr{L}_\mathrm{SM} -\frac{1}{2} \overline{N_\mathrm{R}^i} (M_N)_{ij} N^{c j}_\mathrm{R} -(Y_{N})_{i\alpha}\overline{N_\mathrm{R}^i}  \phi^\dagger
\ell^\alpha_\mathrm{L} +\hc\; .
\end{eqnarray}
Here, $\phi$ denotes the SM Higgs field, which breaks the EW symmetry after acquiring its vev $v_{\mathrm{EW}}$. We have also introduced 
the Majorana mass $M_N$ allowed for the right-handed neutrinos $N_\mathrm{R}^i$ as well as the Yukawa couplings between the neutrinos and the 
Higgs field. We will restrict our study to the extension of the SM by 3 right-handed neutrino fields. The vev of the Higgs will induce Dirac masses 
$m_D = v_\text{EW} Y_N/\sqrt{2}$. Thus, the full $6\times 6$ mixing matrix $U$ is the unitary matrix that diagonalizes the extended neutrino mass matrix: 
\begin{equation}
U^T \left(
\begin{array}{cc}
0 & m_D^T \\ m_D & M_N
\end{array}
\right) U = \left(
\begin{array}{cc}
m & 0 \\ 0 & M
\end{array}
\right), \label{eq:diag}
\end{equation}
where $m$ and $M$ are diagonal matrices containing respectively the masses of the 3 light $\nu_i$ and 3 heavy $N_i$ mass eigenstates. 
The diagonalizing matrix $U$ can be written as~\cite{Blennow:2011vn}:
\begin{equation}
U = \left(
\begin{array}{cc}
\ c & s \\ -s^\dagger & \hat{c}
\end{array}
\right) \left(
\begin{array}{cc}
U_{\rm PMNS} & 0 \\ 0 & I %\mathbb{1}
\end{array}
\right), \label{eq:block}
\end{equation}
where
\begin{equation}
\left(
\begin{array}{cc}
\ c & s \\ -s^\dagger & \hat{c}
\end{array}
\right) \equiv \left(
\begin{array}{cc}
\displaystyle\sum\limits_{n=0}^\infty \frac{ \left(- \Theta \Theta^\dagger \right)^{n}}{(2n)!} & 
\displaystyle\sum\limits_{n=0}^\infty \frac{ \left(- \Theta \Theta^\dagger \right)^{n}}{\left(2n+1\right)!} \Theta  \\ 
-\displaystyle\sum\limits_{n=0}^\infty \frac{ \left(- \Theta^\dagger \Theta \right)^{n}}{\left(2n+1\right)!} \Theta^\dagger & 
\displaystyle\sum\limits_{n=0}^\infty \frac{ \left(- \Theta^\dagger \Theta \right)^{n}}{2n!}\end{array}
\right)
\label{eq:sincos}
\end{equation}
and $U_{\rm PMNS}$ is, \emph{approximately}, the PMNS matrix measured in neutrino oscillation experiments up to the non-Unitary (Hermitian) corrections from $c$. For alternative parametrizations of the full mixing matrix see Refs~\cite{Schechter:1980gr,Schechter:1981cv,Xing:2007zj,Xing:2011ur,Donini:2012tt}. Indeed, due to this Hermitian correction, the actual PMNS matrix appearing in charge current interactions mixing the light neutrinos and charged leptons will, in general, not be Unitary and we will refer to it as $N$: 
\begin{equation}
N = c\,U_{\rm PMNS} 
\label{eq:N}
\end{equation}
The general matrix $\Theta$, representing the mixing between active ($\nu_e$, $\nu_\mu$ and $\nu_\tau$) and heavy ($N_1$, $N_2$ and $N_3$) neutrino states, and the mass eigenstates $m$ and $M$ are determined from Eq.~(\ref{eq:diag}) which leads to:
\begin{equation}
\label{eq:corr}
c^* U_{\rm PMNS} ^* m U_{\rm PMNS} ^\dagger c = - s^* M s^\dagger .
\end{equation}
In the Seesaw limit, that is $M_N \gg m_D$, these conditions reduce to the well-known results:
\begin{eqnarray}
\nonumber
\Theta &\simeq& m_D^\dagger M_N^{-1} \\
\nonumber
U_{\rm PMNS} ^* m U_{\rm PMNS} ^\dagger  &\simeq& - m_D^t M_N^{-1} m_D \equiv -\hat{m}\\
M &\simeq& M_N .
\label{eq:seesaw}
\end{eqnarray}

Notice that, naively, the mixing between the active and heavy neutrinos $\Theta \Theta^\dagger \sim m/M$ and, given the smallness of neutrino masses $m$, the mixing effects we will study here would be unobservably small. However, in the context of Seesaw mechanisms with an approximate conservation of $B-L$ such as the inverse~\cite{Mohapatra:1986bd,Bernabeu:1987gr} or the linear~\cite{Malinsky:2005bi} Seesaws, this symmetry suppresses the neutrino mass $m$ while allowing a sizable mixing. This approximate symmetry not only ensures an equally approximate cancellation in the combination $m_D^t M_N^{-1} m_D$ leading to the observed neutrino masses while allowing large -potentially observable- $\Theta \Theta^\dagger = m_D^\dagger M_N^{-2} m_D$, but also ensures the radiative stability and technical naturalness of the scheme~\cite{Kersten:2007vk}. 

When extending the SM Lagrangian by only 3 new singlet (right-handed neutrino) fields essentially the only neutrino mass matrix with an underlying $L$ symmetry that leads to 3 heavy massive neutrinos is~\cite{Abada:2007ux} (see also Ref.~\cite{Adhikari:2010yt}):
\begin{equation}
m_D = \frac{v_\text{EW}}{\sqrt{2}} \left(
\begin{array}{ccc}
Y_e & Y_\mu & Y_\tau \\ \epsilon_1 Y'_e & \epsilon_1 Y'_\mu & \epsilon_1 Y'_\tau \\ \epsilon_2 Y''_e & \epsilon_2 Y''_\mu & \epsilon_2 Y''_\tau
\end{array}
\right)
\qquad
\textrm{and}
\qquad
M_N = \left(
\begin{array}{ccc}
\mu_1 & \Lambda & \mu_3 \\ \Lambda & \mu_2 & \mu_4 \\ \mu_3 & \mu_4 & \Lambda'
\end{array}
\right), \label{eq:texture}
\end{equation}
with all $\epsilon_i$ and $\mu_j$ small lepton number violating parameters  (see also Ref.~\cite{Dev:2013oxa} for a particular scenario where these small parameters arise naturally). Indeed, setting all $\epsilon_i=0$ and $\mu_j=0$, lepton number symmetry is recovered with the following $L$ assignments $L_e = L_\mu = L_\tau = L_1 = -L_2 = 1$ and $L_3 = 0$. In Eq.~(\ref{eq:seesaw}) this leads to: $\hat{m}=0$ (3 massless neutrinos in the $L$-conserving limit), $M_1=M_2=\Lambda$ (a heavy Dirac pair) and $M_3=\Lambda'$ (a heavy decoupled Majorana singlet), but:
\begin{equation}
\Theta = \frac{v_\text{EW}}{2 \Lambda} \left(
\begin{array}{ccc}
-i Y_e^* & Y_e^* & 0 \\ -i Y_\mu^* & Y_\mu^* & 0 \\ -i Y_\tau^* & Y_\tau^* & 0
\end{array}
\right)
\equiv \frac{1}{\sqrt{2}} \left(
\begin{array}{ccc}
-i \theta_e & \theta_e & 0 \\ -i \theta_\mu & \theta_\mu & 0 \\ -i \theta_\tau & \theta_\tau & 0
\end{array}
\right)
\textrm{and}
\quad
\Theta \Theta^\dagger=  \left(
\begin{array}{ccc}
|\theta_e|^2 & \theta_e \theta_\mu^* & \theta_e \theta_\tau^* \\ \theta_\mu \theta_e^* & |\theta_\mu|^2 & \theta_\mu \theta_\tau^* \\ \theta_\tau \theta_e^* & \theta_\tau \theta_\mu^* & |\theta_\tau|^2
\end{array}
\right). \label{eq:theta}
\end{equation}

Thus, vanishing light neutrino masses can still be associated with arbitrarily large mixing between the heavy Dirac pair and active neutrinos and, for these kind of Seesaw scenarios, the bounds on the mixing we will explore are complementary and independent to the stringent constraints on the absolute light neutrino mass scale. 

The small $L$-violating parameters $\epsilon_i$ and $\mu_j$ will induce small non-zero neutrino masses and mixing among these light mass eigenstates but will only translate in negligible perturbations to the matrix $\Theta$. With the simple form in Eq.~(\ref{eq:theta}) for the heavy-active mixing, the series expansions in Eq.~(\ref{eq:sincos}) can be added exactly obtaining:
\begin{equation}
s= \frac{\sin \theta}{\theta} \Theta
\qquad
\textrm{and}
\qquad
c = I -\frac{1-\cos \theta}{\theta^2} \Theta \Theta^\dagger,
\label{eq:summed}
\end{equation}
with
\begin{equation}
\theta = \sqrt{|\theta_e|^2 + |\theta_\mu|^2 + |\theta_\tau|^2}.
\end{equation}

Regarding the role of the $\epsilon_i$ and $\mu_j$ parameters in the generation of the light neutrino masses and mixings observed in neutrino oscillations, 
all of them except $\mu_1$ and $\mu_3$ will lead to $\hat{m} \neq 0$ through Eq.~(\ref{eq:seesaw}) when switched on:

\begin{eqnarray}
\hat{m}&=&\left( \mu_2 +\frac{\mu_4^2}{\Lambda'}\right)\mathbf{m^{\mathit{t}}_D}\Lambda^{-2}\mathbf{m_D}
-\epsilon_1 \mathbf{m'^{\mathit{t}}_D} \Lambda^{-1} \mathbf{m_D} - \epsilon_1 \mathbf{m^{\mathit{t}}_D} \Lambda^{-1} \mathbf{m'_D} 
+ \epsilon_2^2 \mathbf{m''^{\mathit{t}}_D} \Lambda'^{-1} \mathbf{m''_D}
\nonumber\\
&+&\epsilon_2 \frac{\mu_4}{\Lambda'} 
\left( \mathbf{m^{\mathit{t}}_D} \Lambda^{-1} \mathbf{m''_D}+ \mathbf{m''^{\mathit{t}}_D} \Lambda'^{-1} \mathbf{m_D} \right),
\label{eq:lightmass0}
\end{eqnarray}
with

\begin{equation}
\mathbf{m_D} \equiv \frac{v_\text{EW}}{\sqrt{2}} (Y_e, Y_\mu, Y_\tau) ,
\qquad
\mathbf{m'_D} \equiv \frac{v_\text{EW}}{\sqrt{2}} (Y'_e, Y'_\mu, Y'_\tau) 
\qquad
\textrm{and}
\qquad
\mathbf{m''_D} \equiv \frac{v_\text{EW}}{\sqrt{2}} (Y''_e, Y''_\mu, Y''_\tau). 
\end{equation}
Indeed, even though $\mu_1$ and $\mu_3$ do violate $L$, upon their inclusion the mass matrix in Eq.~(\ref{eq:diag}) does not increase its rank, 
which, in absence of the other $\epsilon_i$ and $\mu_j$, is only 3 and thus 3 massless eigenstates are still 
recovered\footnote{Notice that, even if $\mu_1$ and $\mu_3$ do not induce neutrino masses at tree level, the $L$ symmetry protecting them is 
now broken and loop contributions would appear instead~\cite{LopezPavon:2012zg}.}. The parameters $\mu_2$ and $\mu_4$ do contribute at tree level 
to generate light neutrino masses, however, their effect can be absorbed in a redefinition of the vectors $\mathbf{m'_D}$ and
 $\mathbf{m''_D}$ as follows: 
\begin{equation}
\epsilon_1 \mathbf{m'_D} \rightarrow \epsilon_1 \mathbf{m'_D} - \frac{\mu_2}{2 \Lambda} \mathbf{m_D}
\qquad
\textrm{and}
\qquad
\epsilon_2 \mathbf{m''_D} \rightarrow \epsilon_2 \mathbf{m''_D} - \frac{\mu_4}{\Lambda} \mathbf{m_D} 
\end{equation} 
up to contributions with two extra powers of the small $L$-violating parameters. Thus, in presence of non-zero $\epsilon_i$, 
it is enough to consider their contribution to the generation of neutrino masses which reads:
\begin{equation}
\hat{m} = \epsilon_1 \mathbf{m'^{\mathit{t}}_D} \Lambda^{-1} \mathbf{m_D} + \epsilon_1 \mathbf{m^{\mathit{t}}_D} \Lambda^{-1} \mathbf{m'_D} + \epsilon_2^2 \mathbf{m''^{\mathit{t}}_D} \Lambda'^{-1} \mathbf{m''_D} .
\label{eq:lightmass}
\end{equation}
Notice that the last term in Eq.~(\ref{eq:lightmass}) is suppressed by two powers of $\epsilon_2$ while the others only by one power of $\epsilon_1$. However, $\epsilon_2$ (and $\mu_3$ and $\mu_4$) violates $L$ by one unit while $\epsilon_1$ (and $\mu_1$ and $\mu_2$) by 2. Hence, if the source of $L$-violation is by one unit it is expected that $\epsilon_1 \sim \epsilon_2^2$. Thus, for full generality, we will keep the last term in Eq.~(\ref{eq:lightmass}). The six free parameters encoded in $\mathbf{m'_D}$ and $\mathbf{m''_D}$ allow to give mass to the three mass eigenstates observed in neutrino oscillations as well as the possibility of reproducing any mixing pattern including the, yet unknown, CP-violating phases of Dirac and Majorana types encoded in the PMNS matrix, while leaving $\mathbf{m_D}$, and hence $\Theta$, $s$ and $c$, mostly unconstrained~\footnote{In contrast, neglecting the last term in Eq.~(\ref{eq:lightmass}) would lead to the more constrained scenario explored in detail in Ref.~\cite{Gavela:2009cd}, with a massless neutrino and a mixing pattern in $\Theta$, $s$ and $c$ determined up to an overall factor from the observed neutrino oscillation parameters. This scenario has also been studied in Refs.~\cite{Zhang:2009ac,Malinsky:2009df,Ibarra:2010xw,Ibarra:2011xn,Cely:2012bz}}. One of the three elements of $\mathbf{m_D}$ is, however, fixed by the other two, the values of the light mass eigenstates and the elements of the PMNS matrix when solving for Eq.~(\ref{eq:lightmass}) obtaining the following relation:
\begin{equation}
\begin{split}
Y_\tau& \simeq \frac{1}{\hat{m}_{e \mu}^2 - \hat{m}_{ee} \hat{m}_{\mu \mu}}\left(Y_e\left(\hat{m}_{e \mu}\hat{m}_{\mu \tau}-\hat{m}_{e \tau}\hat{m}_{\mu \mu}\right)+\right.\\
&\left. Y_\mu\left(\hat{m}_{e \mu}\hat{m}_{e \tau}-\hat{m}_{ee}\hat{m}_{\mu \tau}\right)-\sqrt{Y_e^2\hat{m}_{\mu \mu}-2Y_eY_\mu \hat{m}_{e \mu}+Y_\mu^2\hat{m}_{ee}}\times \right.\\
&\left.\times\sqrt{\hat{m}_{e \tau}^2\hat{m}_{\mu \mu}-2\hat{m}_{e \mu}\hat{m}_{e \tau}\hat{m}_{\mu \tau}+\hat{m}_{ee}\hat{m}_{\mu \tau}^2+\hat{m}_{e \mu}^2\hat{m}_{\tau \tau}-\hat{m}_{ee}\hat{m}_{\mu \mu}\hat{m}_{\tau \tau}}\right) ,
\end{split}
\label{eq:Yt}
\end{equation}
where $\hat{m} = -U_{\rm PMNS} ^* m U_{\rm PMNS} ^\dagger$ is the mass matrix of the flavour eigenstates. Thus, in our numerical exploration of the parameter space in Section~\ref{sec:res} we will consider the 9 free parameters summarized in Table~\ref{tab:params}.  

An alternative parametrization extensively used in the literature is the so-called Casas-Ibarra parametrization~\cite{Casas:2001sr}. This parametrization introduces the matrix $R = i M^{-1/2} m_D U_{\rm PMNS} m^{-1/2}$ exploiting the fact that, from \eq~(\ref{eq:seesaw}), $R$ has to be (complex) orthogonal. The main advantage of this parametrization is the ability to easily recover the Yukawa couplings through the heavy mass eigenvalues $M$ and the low energy observables $U_{\rm PMNS} $ and $m$ together with the elements of $R$ as $m_D = -i M^{1/2} R m^{1/2} U_{\rm PMNS} ^\dagger$. However, the physical range of the parameters contained in $R$ can be cumbersome and a physical interpretation of their values is not immediately transparent, see \Ref~\cite{Casas:2010wm} for a detailed discussion. Moreover, these relations only hold at tree level\footnote{See Ref.~\cite{Lopez-Pavon:2015cga} for a generalization of the Casas-Ibarra approach to loop level.}. Thus, when values of $R$ are chosen so as to allow sizable low energy phenomenology through large Yukawas and low $M$, it is important to check if the pattern displays an approximate $B-L$ symmetry. Otherwise, loop corrections to the unprotected Weinberg operator, that is to $U_{\rm PMNS} $ and $m$, will exceed present constraints even if their values were correct at tree level. For this reason we rather chose to perform the scan through the parameters summarized in Table~\ref{tab:params}.  

\begin{table}
\begin{center}
\begin{tabular}{|c||c|c|c|c|c|c|}
\hline
Parameter & $\left|Y_e\right| \times \left|Y_\mu\right| $ & $\left|Y_e\right| - \left|Y_\mu\right| $ & $m_1$ [eV] &  $\Lambda$ [GeV] & Phases: $\alpha_e$, $\alpha_\mu$, $\delta$, $\alpha_1$ \& $\alpha_2$ & Osc. data \rule{0pt}{2.6ex} \rule[-1.2ex]{0pt}{0pt}\\
\hline
Range &  $(0,10^{-4})$ & $(-0.1,0.1)$ & $(10^{-5},1)$ & $(10^{3},10^{4})$ & $(0,2\pi)$ & fixed \cite{Gonzalez-Garcia:2014bfa} \rule{0pt}{2.6ex} \rule[-1.2ex]{0pt}{0pt} \\
\hline
\end{tabular}
\caption{The 9 free parameters of our scan: the modulus and phase of the electron and muon Yukawas $|Y_e|$, $|Y_\mu|$, $\alpha_e$ and $\alpha_\mu$, the Majorana mass scale $\Lambda$, the absolute neutrino mass $m_1$ and the 3 yet unknown CP-violation phases (Dirac and Majorana) in the PMNS mixing matrix: $\delta$, $\alpha_1$ and $\alpha_2$. The PMNS mixing angles and mass splittings are fixed to their best fit from the global analysis in Ref.~\cite{Gonzalez-Garcia:2014bfa}. }
\label{tab:params}
\end{center}
\end{table}

At energies much below the masses of the heavy neutrinos $\Lambda$ and $\Lambda'$ the effects of their mixing $\Theta$ manifest dominantly through deviations from unitarity of the lepton mixing matrix $N$. Since any general matrix can be parametrized as the product of an Hermitian and a Unitary matrix, these deviations from unitarity have been often parametrized as~\cite{FernandezMartinez:2007ms}:
\begin{equation} 
N = (I - \eta) U_{\rm PMNS}
\end{equation} 
where the small Hermitian matrix $\eta$ (also called $\epsilon$ in other works) corresponds to the coefficient of the only dimension 6 operator obtained at tree level upon integrating out the heavy right-handed neutrinos in a Seesaw scenario~\cite{Broncano:2002rw} and, in our parametrization it would be given from Eqs.~(\ref{eq:N}) and (\ref{eq:summed}) by: 
\begin{equation}
\eta = \frac{1-\cos \theta}{\theta^2} \Theta \Theta^\dagger .
\end{equation}

\section{Observables}
\label{sec:obs}

In this section we introduce the list of observables used for our analysis. While a more comprehensive set could be considered (see for example Ref.~\cite{Antusch:2014woa}), we have rather chosen the most representative of these observables since extending the analysis to the loop level for the whole set would be cumbersome and the dominant constraints as well as the main effects pointed out in~\cite{Akhmedov:2013hec} are contained in a smaller subset. We will thus present both the 1-loop contributions and the experimental constraints for a total of 13 observables. The loop amplitudes of the processes have been computed exploiting the Goldstone-boson equivalence theorem~\cite{Cornwall:1974km} under the assumption that the mass of the extra neutrinos $M_i$ is larger than the gauge boson masses; i.e. $M_i > M_{W,Z}$. Thus, we have made the simplifying assumption that the most relevant loop corrections are those were the loops are mediated by either the Higgs boson, $h$, the Goldstone bosons $\phi^\pm$ and $\phi^0$ or the heavy Majorana neutrinos. Indeed, this forces the vertexes to involve the potentially large Yukawa couplings (the only couplings that can be relevant at the loop level) and the corrections from including the transverse components are suppressed by $M_{W,Z}^2/M_N^2$. The set of 13 independent observables analyzed in this study is composed of:

\begin{itemize}
\item 8 ratios constraining electroweak universality: $R^\pi_{\mu e}$, $R^\pi_{\tau \mu}$, $R^W_{\mu e}$, $R^W_{\tau \mu}$, $R^K_{\mu e}$, $R^K_{\tau \mu}$, $R^l_{\mu e}$, $R^l_{\tau \mu}$
\item The invisible $Z$ width
\item The $W$ mass $M_W$
\item 3 rare flavour-changing decays: $\mu\rightarrow e\gamma$, $\tau\rightarrow \mu\gamma$ and $\tau\rightarrow e\gamma$
\end{itemize}
All of them will be determined as a function of the three most precise electroweak measurements: $\alpha$, $M_Z$ and $G_\mu$ ($G_F$ as measured from $\mu$ decay)~\cite{Agashe:2014kda}:
\begin{eqnarray}
\alpha&=&\left(7.2973525698\pm0.0000000024\right)\times 10^{-3}, \nonumber\\
M_Z&=&\left(91.1876\pm0.0021\right) \text{ GeV}, \\
G_\mu&=&\left(1.1663787\pm0.0000006\right)\times 10^{-5} \text{ GeV}^{-2}. \nonumber
\end{eqnarray}

\begin{figure}
\centering
\begin{picture}(550,60)
\Photon(70,0)(110,0){5}{3}
\put(120,0){$=$}
\Photon(140,0)(180,0){5}{3}
\put(190,0){$+$}
\Photon(210,0)(250,0){5}{3}
\Arc(270,0)(20,0,360)
\Photon(290,0)(340,0){5}{3}
\put(85,12){$W$}
\put(155,12){$W$}
\put(225,12){$W$}
\put(265,-33){$N$}
\put(270,25){$l$}
\put(305,12){$W$}
\end{picture}

\vspace{1cm}

\begin{picture}(550,60)
\Photon(70,0)(110,0){5}{3}
\put(120,0){$=$}
\Photon(140,0)(180,0){5}{3}
\put(190,0){$+$}
\Photon(210,0)(250,0){5}{3}
\Arc(270,0)(20,0,360)
\Photon(290,0)(340,0){5}{3}
\put(85,12){$Z$}
\put(155,12){$Z$}
\put(225,12){$Z$}
\put(265,-33){$N$}
\put(265,25){$N$}
\put(305,12){$Z$}
\end{picture}

\vspace{1cm}

\caption{1-loop correction of the new heavy neutrinos to $W$ and $Z$ propagators.}
\label{fig:propa}
\end{figure}
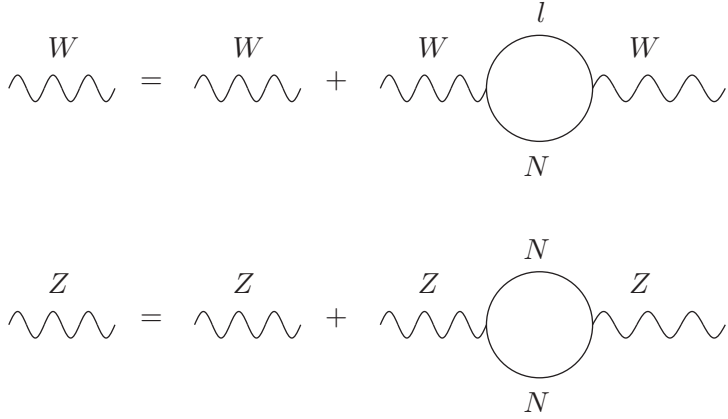

All observables will receive contributions from the loop corrections to the $W$ and $Z$ boson propagators through the diagrams in Fig.~\ref{fig:propa}. These contributions are encoded in the flavour-universal corrections $\delta_{W,Z}^\text{univ}$ that can be found in Eq.~(\ref{dW}) in the Appendix. We now list the further corrections exclusive to each of the observables considered:

\subsection{$\mu$ decay, $G_F$ and $M_W$}

Our input value for $G_F$ is determined through $\mu$ decay, but this process will receive corrections both at the tree and the loop level (see Fig.~\ref{fig:GF}). Thus, the value determined from $\mu$ decay, $G_\mu$, is related to $G_F$ by:

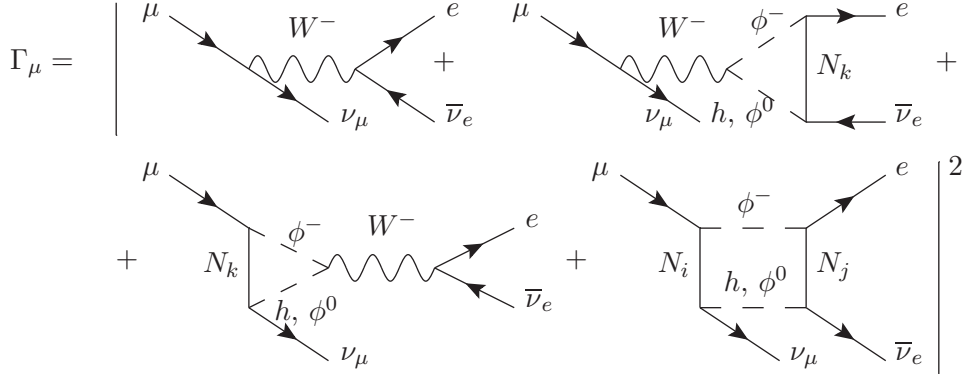
\begin{figure}
\centering
\begin{picture}(550,40)
\put(0,0){$\Gamma_{\mu}=$}
\Line(40,-25)(40,25)
\Line[arrow,arrowpos=0.5,arrowlength=5,arrowwidth=2,arrowinset=0.2](60,20)(90,0)
\Line[arrow,arrowpos=0.5,arrowlength=5,arrowwidth=2,arrowinset=0.2](90,0)(120,-20)
\Photon(90,0)(130,0){5}{3}
\Line[arrow,arrowpos=0.5,arrowlength=5,arrowwidth=2,arrowinset=0.2](130,0)(160,20)
\Line[arrow,arrowpos=0.5,arrowlength=5,arrowwidth=2,arrowinset=0.2](160,-20)(130,0)
\put(160,0){$+$}
\Line[arrow,arrowpos=0.5,arrowlength=5,arrowwidth=2,arrowinset=0.2](200,20)(230,0)
\Line[arrow,arrowpos=0.5,arrowlength=5,arrowwidth=2,arrowinset=0.2](230,0)(260,-20)
\Photon(230,0)(270,0){5}{3}
\Line[dash,dashsize=8](270,0)(300,20)
\Line[dash,dashsize=8](300,-20)(270,0)
\Line(300,20)(300,-20)
\Line[arrow,arrowpos=0.5,arrowlength=5,arrowwidth=2,arrowinset=0.2](300,20)(330,20)
\Line[arrow,arrowpos=0.5,arrowlength=5,arrowwidth=2,arrowinset=0.2](330,-20)(300,-20)
\put(350,0){$+$}
\put(40,-75){$+$}
\Line[arrow,arrowpos=0.5,arrowlength=5,arrowwidth=2,arrowinset=0.2](60,-40)(90,-60)
\Line(90,-60)(90,-90)
\Line[arrow,arrowpos=0.5,arrowlength=5,arrowwidth=2,arrowinset=0.2](90,-90)(120,-110)
\Line[dash,dashsize=8](90,-60)(120,-75)
\Line[dash,dashsize=8](90,-90)(120,-75)
\Photon(120,-75)(160,-75){5}{3}
\Line[arrow,arrowpos=0.5,arrowlength=5,arrowwidth=2,arrowinset=0.2](160,-75)(190,-60)
\Line[arrow,arrowpos=0.5,arrowlength=5,arrowwidth=2,arrowinset=0.2](190,-90)(160,-75)
\put(210,-75){$+$}
\Line[arrow,arrowpos=0.5,arrowlength=5,arrowwidth=2,arrowinset=0.2](230,-40)(260,-60)
\Line(260,-60)(260,-90)
\Line[arrow,arrowpos=0.5,arrowlength=5,arrowwidth=2,arrowinset=0.2](260,-90)(290,-110)
\Line[dash,dashsize=8](260,-90)(300,-90)
\Line[dash,dashsize=8](260,-60)(300,-60)
\Line(300,-60)(300,-90)
\Line[arrow,arrowpos=0.5,arrowlength=5,arrowwidth=2,arrowinset=0.2](300,-60)(330,-40)
\Line[arrow,arrowpos=0.5,arrowlength=5,arrowwidth=2,arrowinset=0.2](300,-90)(330,-110)
\Line(350,-35)(350,-115)
\put(50,20){$\mu$}
\put(190,20){$\mu$}
\put(50,-40){$\mu$}
\put(220,-40){$\mu$}
\put(125,-20){$\nu_\mu$}
\put(240,-20){$\nu_\mu$}
\put(125,-110){$\nu_\mu$}
\put(295,-110){$\nu_\mu$}
\put(105,12){$W^-$}
\put(245,12){$W^-$}
\put(135,-63){$W^-$}
\put(280,15){$\phi^-$}
\put(265,-20){$h,\,\phi^0$}
\put(105,-65){$\phi^-$}
\put(100,-95){$h,\,\phi^0$}
\put(275,-54){$\phi^-$}
\put(270,-84){$h,\,\phi^0$}
\put(305,-2){$N_k$}
\put(73,-77){$N_k$}
\put(245,-77){$N_i$}
\put(305,-77){$N_j$}
\put(165,20){$e$}
\put(165,-20){$\overline{\nu}_e$}
\put(335,20){$e$}
\put(335,-20){$\overline{\nu}_e$}
\put(195,-60){$e$}
\put(195,-90){$\overline{\nu}_e$}
\put(335,-40){$e$}
\put(335,-110){$\overline{\nu}_e$}
\put(355,-40){$2$}
\end{picture}
\vspace{3cm}
\caption{1-loop corrections to $\mu$ decay.}
\label{fig:GF}
\end{figure}

\begin{equation}
\Gamma_\mu = \frac{m_\mu^5 G_F^2}{192 \pi^3}\left(1 - |\theta_e|^2 - |\theta_\mu|^2 + 2\delta^{\text{univ } N}_W+\delta G\right)
\equiv \frac{m_\mu^5 G_\mu^2}{192 \pi^3} ,
\label{mudecay}
\end{equation}
with

\begin{equation}
\delta G = 2Re[\mathcal{V}^W_e+\mathcal{V}^{W *}_\mu+\delta^{\text{CT } W}_{e}+\delta^{\text{CT } W *}_{\mu}+\mathcal{B}_{\mu e}]
\end{equation}
and where $\delta^{\text{univ } N}_W$ is the flavour-universal $W$ propagator correction, $\delta^{\text{CT } W}_{l}$ and $\mathcal{V}^W_{l}$ are the flavour-dependent lepton propagator and vertex contributions (see Eqs.~(\ref{eq:propscorr}) and~(\ref{eq:Wvertex}) in the Appendix), and $\mathcal{B}_{\mu e}$ encodes the box diagram contribution computed in Eq.~(\ref{eq:box}) in the Appendix.  

From Eq.~(\ref{mudecay}), we find:
\begin{equation}
G_\mu^2 = G_F^2 \left( 1 - |\theta_e|^2 - |\theta_\mu|^2 + 2\delta_W^\text{univ N}+\delta G\right)\,.
\label{Gmu}
\end{equation}

The second and third terms in Eq.~(\ref{Gmu}) correspond to the tree level correction, the fourth term is the universal 1-loop oblique correction 
which is given in Eq.~(\ref{dW}) of the Appendix. This particular expression, when used in an observable mediated by the $Z$ and thus corrected 
through $2\delta_Z^\text{univ N}$, leads to a common correction to these observables given by $1 - |\theta_e|^2 - |\theta_\mu|^2 - 2 \alpha T$ 
(see Eqs.~(\ref{dW}) and (\ref{T})). This common dependence on the tree level and oblique corrections is the source of the cancellation analyzed in Ref.~\cite{Akhmedov:2013hec}.

The the $W$ mass is also correlated to $G_F$ through
\begin{equation}
M_W^2 = \frac{\pi \alpha}{\sqrt{2} G_F s_\mathrm{W}^2 (1 - \Delta r)} , 
\end{equation}
with $\Delta r = 0.03639 \mp 0.00036 \pm 0.00011$~\cite{Agashe:2014kda}. Thus, the corrections induced at both the tree and loop levels by 
the heavy neutrinos from Eq.~(\ref{Gmu}) can be probed by the measurement of $M_W$ in LEP and Tevatron~\cite{Agashe:2014kda}:
\begin{equation}
M_W = 80.385 \pm 0.015 \quad \mathrm{GeV} .
\end{equation}

\subsection{Invisible $Z$ width}

The determination of the number of light active neutrinos by LEP through the invisible width of the $Z$ provides a constraint to heavy neutrino mixing already at the tree level. Additional loop corrections are induced through the diagrams in Fig.~\ref{fig:invZ} which lead to:
\begin{equation}
\Gamma_\text{inv}=\displaystyle\sum_{i,j=1}^3 \frac{G_F M_Z^3 \rho}{24 \sqrt{2} \pi}\left( \mathcal{Z}_{ij}+\mathcal{Z}_{ji}\right)\, ,
\label{eq:inv}
\end{equation}
where $\rho$ encodes the SM loop corrections to the process and 
\begin{equation}
\mathcal{Z}_{ij}=\vert C_{ij}\vert ^2\big(1+\delta_Z^\text{univ}\big)+2 Re\big[C_{ij}^*\left(\delta^{\text{CT } Z}_{ij}+\mathcal{V}^Z_{ij}\right)\big]\, ,
\end{equation}
with
\begin{equation}
C_{ij}=\sum_{\alpha=e,\mu,\tau} U_{\alpha i}^* U_{\alpha j}\,.
\end{equation}
and $\delta^{\text{CT } Z}_{ij}$ and $\mathcal{V}^Z_{ij}$ the lepton and vertex corrections shown in Eqs.~(\ref{eq:propscorrZ}) and~(\ref{eq:Zvertex}) 
in the Appendix. 

Eq.~(\ref{eq:inv}) is often used to determine the number of active neutrinos $N_\nu$ lighter than $M_Z/2$ as:
\begin{equation}
\Gamma_\text{inv}=  \frac{G_F M_Z^3 \rho N_\nu}{12 \sqrt{2} \pi}\, ,
\label{eq:invi}
\end{equation}
The measurement by LEP of $\Gamma_\text{inv}=\left(0.4990\pm0.0015\right) \text{ GeV}$ combined with Eq.~(\ref{eq:invi}) leads to~\cite{Agashe:2014kda}:
\begin{equation}
N_\nu = 2.990 \pm 0.007 \, .
\end{equation}
We will exploit this result together with Eq.~(\ref{eq:inv}) to derive constraints on $C_{ij}$ and, hence, on the heavy neutrino mixings.

\begin{figure}
\centering
\begin{picture}(550,40)
\put(0,0){$\Gamma_\text{inv}=$}
\Line(40,-25)(40,25)
\Photon(50,0)(90,0){5}{3}
\Line[arrow,arrowpos=0.5,arrowlength=5,arrowwidth=2,arrowinset=0.2](90,0)(130,20)
\Line[arrow,arrowpos=0.5,arrowlength=5,arrowwidth=2,arrowinset=0.2](130,-20)(90,0)
\put(130,0){$+$}
\Photon(150,0)(190,0){5}{3}
\Line(230,-20)(230,20)
\Line[dash,dashsize=8](190,0)(230,20)
\Line[dash,dashsize=8](190,0)(230,-20)
\Line[arrow,arrowpos=0.5,arrowlength=5,arrowwidth=2,arrowinset=0.2](230,20)(270,20)
\Line[arrow,arrowpos=0.5,arrowlength=5,arrowwidth=2,arrowinset=0.2](270,-20)(230,-20)
\put(270,0){$+$}
\Photon(290,0)(330,0){5}{3}
\Line[dash,dashsize=8](370,-20)(370,20)
\Line(330,0)(370,20)
\Line(370,-20)(330,0)
\Line[arrow,arrowpos=0.5,arrowlength=5,arrowwidth=2,arrowinset=0.2](370,20)(410,20)
\Line[arrow,arrowpos=0.5,arrowlength=5,arrowwidth=2,arrowinset=0.2](410,-20)(370,-20)
\Line(430,-25)(430,25)
\put(70,12){$Z$}
\put(170,12){$Z$}
\put(310,12){$Z$}
\put(205,-22){$h$}
\put(200,16){$\phi^0$}
\put(235,-2){$N_k$}
\put(135,-20){$\overline{\nu}_i$}
\put(135,20){$\nu_j$}
\put(275,-20){$\overline{\nu}_i$}
\put(275,20){$\nu_j$}
\put(415,-20){$\overline{\nu}_i$}
\put(415,20){$\nu_j$}
\put(375,-2){$h,\, \phi^0$}
\put(345,-24){$N_a$}
\put(345,20){$N_b$}
\put(435,20){$2$}
\end{picture}
\vspace{1cm}
\caption{1-loop corrections to the invisible decay of the $Z$.}
\label{fig:invZ}
\end{figure}
 
%%%%%%%%

\subsection{Universality ratios}

Electroweak coupling universality is strongly constrained through ratios of leptonic decays of $K$, $\pi$, $W$ or charged leptons. In these ratios many uncertainties cancel and a clean constraint can be derived. These observables are corrected both at the tree and loop level, for instance, $R^\pi_{\mu e}=\Gamma\left(\pi^- \rightarrow \mu \overline{\nu}_\mu\right)/\Gamma\left(\pi^- \rightarrow e \overline{\nu}_e\right)$ is corrected by the diagrams in Fig.~\ref{fig:univ}.

\begin{figure}
\centering
\begin{picture}(550,40)
\put(0,-32){$R^\pi_{\mu e}=$}
\Line(35,-30)(420,-30)
\Line(40,-25)(40,25)
\Line[arrow,arrowpos=0.5,arrowlength=5,arrowwidth=2,arrowinset=0.2](60,20)(100,0)
\Line[arrow,arrowpos=0.5,arrowlength=5,arrowwidth=2,arrowinset=0.2](100,0)(60,-20)
\Photon(100,0)(140,0){5}{3}
\Line[arrow,arrowpos=0.5,arrowlength=5,arrowwidth=2,arrowinset=0.2](140,0)(180,20)
\Line[arrow,arrowpos=0.5,arrowlength=5,arrowwidth=2,arrowinset=0.2](180,-20)(140,0)
\put(200,0){$+$}
\Line[arrow,arrowpos=0.5,arrowlength=5,arrowwidth=2,arrowinset=0.2](230,20)(270,0)
\Line[arrow,arrowpos=0.5,arrowlength=5,arrowwidth=2,arrowinset=0.2](270,0)(230,-20)
\Photon(270,0)(310,0){5}{3}
\Line[dash,dashsize=8](310,0)(350,20)
\Line[dash,dashsize=8](350,-20)(310,0)
\Line(350,20)(350,-20)
\Line[arrow,arrowpos=0.5,arrowlength=5,arrowwidth=2,arrowinset=0.2](350,20)(390,20)
\Line[arrow,arrowpos=0.5,arrowlength=5,arrowwidth=2,arrowinset=0.2](390,-20)(350,-20)
\put(50,-20){$\overline{u}$}
\put(50,15){$d$}
\put(117,12){$W^-$}
\put(185,15){$\mu$}
\put(185,-20){$\overline{\nu}_\mu$}
\put(220,-20){$\overline{u}$}
\put(220,15){$d$}
\put(395,15){$\mu$}
\put(395,-20){$\overline{\nu}_\mu$}
\put(287,12){$W^-$}
\put(325,16){$\phi^-$}
\put(315,-22){$h,\, \phi^0$}
\put(355,-2){$N_k$}
\Line(410,25)(410,-25)
\put(415,20){$2$}
\Line(40,-35)(40,-85)
\Line[arrow,arrowpos=0.5,arrowlength=5,arrowwidth=2,arrowinset=0.2](60,-40)(100,-60)
\Line[arrow,arrowpos=0.5,arrowlength=5,arrowwidth=2,arrowinset=0.2](100,-60)(60,-80)
\Photon(100,-60)(140,-60){5}{3}
\Line[arrow,arrowpos=0.5,arrowlength=5,arrowwidth=2,arrowinset=0.2](140,-60)(180,-40)
\Line[arrow,arrowpos=0.5,arrowlength=5,arrowwidth=2,arrowinset=0.2](180,-80)(140,-60)
\put(200,-60){$+$}
\Line[arrow,arrowpos=0.5,arrowlength=5,arrowwidth=2,arrowinset=0.2](230,-40)(270,-60)
\Line[arrow,arrowpos=0.5,arrowlength=5,arrowwidth=2,arrowinset=0.2](270,-60)(230,-80)
\Photon(270,-60)(310,-60){5}{3}
\Line[dash,dashsize=8](310,-60)(350,-40)
\Line[dash,dashsize=8](350,-80)(310,-60)
\Line(350,-40)(350,-80)
\Line[arrow,arrowpos=0.5,arrowlength=5,arrowwidth=2,arrowinset=0.2](350,-40)(390,-40)
\Line[arrow,arrowpos=0.5,arrowlength=5,arrowwidth=2,arrowinset=0.2](390,-80)(350,-80)
\put(50,-80){$\overline{u}$}
\put(50,-45){$d$}
\put(117,-48){$W^-$}
\put(185,-45){$e$}
\put(185,-80){$\overline{\nu}_e$}
\put(220,-80){$\overline{u}$}
\put(220,-45){$d$}
\put(395,-45){$e$}
\put(395,-80){$\overline{\nu}_e$}
\put(287,-48){$W^-$}
\put(325,-44){$\phi^-$}
\put(315,-82){$h,\, \phi^0$}
\put(355,-62){$N_k$}
\Line(410,-35)(410,-85)
\put(415,-40){$2$}
\end{picture}
\vspace{2cm}
\caption{1-loop corrections to weak universality ratios.}
\label{fig:univ}
\end{figure}

\begin{table}
\begin{center}
\begin{tabular}{c|c}
\hline
$BR\left(\pi^+\rightarrow e^+\nu_e\right)$& $\left(1.230\pm 0.004\right)\times 10^{-4}$\\
$BR\left(\pi^+\rightarrow \mu^+\nu_\mu\right)$& $\left(99.98770\pm 0.00004\right)\%$\\
$BR\left(\tau^-\rightarrow \pi^-\nu_\tau\right)$& $\left(10.83\pm 0.06\right)\%$\\
$BR\left(K^+\rightarrow e^+\nu_e\right)$& $\left(1.581\pm 0.008\right)\times 10^{-5}$\\
$BR\left(K^+\rightarrow \mu^+\nu_\mu\right)$& $\left(63.55\pm 0.11\right)\% 10^{-5}$\\
$BR\left(\tau^-\rightarrow K^-\nu_\tau\right)$& $\left(7.00\pm 0.10\right)\times 10^{-3}$\\
$BR\left(W^+\rightarrow e^+\nu_e\right)$& $\left(10.71\pm 0.16\right)\%$\\
$BR\left(W^+\rightarrow \mu^+\nu_\mu\right)$& $\left(10.63\pm 0.15\right)\%$\\
$BR\left(W^+\rightarrow \tau^+\nu_\tau\right)$& $\left(11.38\pm 0.21\right)\%$\\
$BR\left(\tau^-\rightarrow \mu^-\overline{\nu}_\mu \nu_\tau\right)$& $\left(17.41\pm 0.04\right)\%$\\
$BR\left(\tau^-\rightarrow e^-\overline{\nu}_e \nu_\tau\right)$& $\left(17.83\pm 0.04\right)\%$\\
\hline
$\tau_{\pi^\pm}$ & $\left(2.6033\pm 0.0005\right)\times 10^{-8}\text{ s}$\\
$\tau_{K^\pm}$ & $\left(1.2380\pm 0.0021\right)\times 10^{-8}\text{ s}$\\
$\tau_\tau$ & $\left(290.3\pm 5.0\right)\times 10^{-15}\text{ s}$\\
$\tau_\mu$ & $\left(2.1969811\pm 0.0000022\right)\times 10^{-6}\text{ s}$\\
\hline
$m_{\pi^\pm}$ & $139.57018\pm 0.00035 \text{ MeV}$ \\
$m_{K^\pm}$ & $493.677\pm 0.016 \text{ MeV}$ \\
$M_{W}$ & $80.385\pm 0.0015 \text{ MeV}$ \\
$m_{e}$ & $0.510998928\pm 0.000000011 \text{ MeV}$ \\
$m_{\mu}$ & $105.6583715\pm 0.0000035 \text{ MeV}$ \\
$m_{\tau}$ & $1776.82\pm 0.16 \text{ MeV}$ \\
\hline
$\delta R^\pi_{\mu e}$ & $\left(-0.374\pm 0.001\right)$ \\
$\delta R^\pi_{\mu \tau}$ & $\left(0.0016\pm 0.0014\right)$ \\
$\delta R^K_{\mu \tau}$ & $\left(0.0090\pm 0.0022\right)$ \\
\hline
\end{tabular}
\caption{Input values used for the constraints on weak universality from ratios of meson and charged lepton decays.}
\label{tab:exp-values}
\end{center}
\end{table}

Thus, the general expression for the ratio of lepton flavours $\alpha$ and $\beta$ is given by:

\begin{equation}
R_{\alpha \beta}=R^{SM}_{\alpha \beta}\frac{\displaystyle1 - |\theta_\alpha|^2 +2\text{Re}\left[\mathcal{V}^W_{\alpha}+\delta^{\text{CT }W}_{\alpha}\right]}{\displaystyle1 - |\theta_\beta|^2 +2\text{Re}\left[\mathcal{V}^W_{\beta}+\delta^{\text{CT }W}_{\beta}\right]} ,
\label{eq:univ1}
\end{equation}
where $R^{SM}_{\alpha \beta}$ is the SM value for this ratio, for example, for $\pi$ decay:
\begin{equation}
R^{\pi SM}_{\alpha \beta} = \left(\frac{ m_\alpha \left(m_\pi^2-m_\alpha^2\right)}{m_\beta\left(m_\pi^2-m_\beta^2\right)}\right)^2\frac{1}{1+\delta R_{\alpha \beta}^\pi}
\label{eq:univ2}
\end{equation}
and where $\delta R_{\alpha \beta}^\pi$ are the SM radiative corrections to this process~\cite{Decker:1994ea}. Notice that the flavour-universal contributions from the $W$ propagator cancel in the ratio. 

The predicted values of these ratios are computed through Eqs.~(\ref{eq:univ1}) and~(\ref{eq:univ2}) with data form \cite{Agashe:2014kda, Pich:2013lsa} and compared to the experimental measurements of the decay rates in our global fit. This data is summarized in Table \ref{tab:exp-values}.

\subsection{Rare decays}

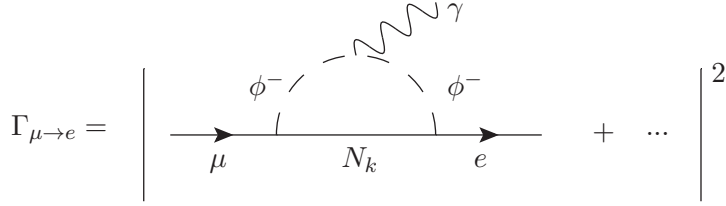
\begin{figure}
\centering
\begin{picture}(550,50)
\put(0,0){$\Gamma_{\mu\rightarrow e} =$}
\Line(50,25)(50,-25)
\Line[arrow,arrowpos=0.5,arrowlength=5,arrowwidth=2,arrowinset=0.2](60,0)(100,0)
\Arc[dash,dashsize=8](130,0)(30,0,-180)
\Line(100,0)(160,0)
\Line[arrow,arrowpos=0.5,arrowlength=5,arrowwidth=2,arrowinset=0.2](160,0)(200,0)
\Photon(130,30)(160,50){5}{3}
\put(75,-12){$\mu$}
\put(175,-12){$e$}
\put(125,-12){$N_k$}
\put(90,15){$\phi^-$}
\put(165,15){$\phi^-$} 
\put(165,45){$\gamma$} 
\put(220,-2){$+$} 
\put(240,0){$...$}
\Line(260,25)(260,-25)
\put(265,20){$2$}
\end{picture}
\vspace{0.5cm}
\caption{Extra neutrino contributions to the $\mu \to e \gamma$ decay.}
\label{fig:muegamma}
\end{figure}

The presence of extra heavy neutrinos beyond the three light ones participating in low energy weak processes induces deviations from unitarity in the PMNS matrix. Thus, the GIM cancellation~\cite{Glashow:1970gm} suppressing flavour-changing processes does not take place and strong constraints on the presence of these extra neutrinos can be derived. Moreover, the extra heavy neutrinos themselves also mediate the flavour-changing processes, such as radiative leptons decays $l_\alpha \to l_\beta \gamma$ in Fig.~\ref{fig:muegamma}. The contribution from both the heavy and light neutrinos is given by:
\begin{equation}
\frac{\Gamma\left(l_\alpha\rightarrow l_\beta \gamma \right)}{\Gamma\left(l_\alpha\rightarrow l_\beta \nu_\alpha \overline{\nu}_\beta \right)}=\frac{3 \alpha}{32 \pi}\frac{\Big| \displaystyle\sum_{k=1}^6 U_{\alpha k}U^\dagger_{k\beta}F(x_k)\Big|^2}{\left(U U^\dagger \right)_{\alpha\alpha}\left(U U^\dagger \right)_{\beta\beta}}
\label{eq:raredecaye}
\end{equation}
where $x_k\equiv \frac{M_{k}^2}{M_W^2}$, and $F(x_k)$ is given by:
\begin{equation}
F(x_k)\equiv\frac{10-43x_k+78x_k^2-49x_k^3+4x_k^4+18x_k^3\ln x_k}{3(x_k-1)^4}.
\end{equation}
Thus, for heavy neutrino masses much larger than $M_W$:
\begin{equation}
\frac{\Gamma\left(l_\alpha\rightarrow l_\beta \gamma \right)}{\Gamma\left(l_\alpha\rightarrow l_\beta \nu_\alpha \overline{\nu}_\beta \right)} \simeq \frac{3 \alpha}{32 \pi} |\theta_\alpha \theta^*_\beta|^2 (F(\infty) -F(0))^2 .
\label{eq:raredecay}
\end{equation}
The prediction from Eq.~(\ref{eq:raredecaye}) will be compared with the existing upper bounds from \cite{Agashe:2014kda}:
\begin{eqnarray}
BR_{\mu e}&<& 5.7\times 10^{-13}\, ,\\ 
BR_{\tau e}&< &3.3\times 10^{-8}\, ,\\
BR_{\tau \mu}&<&4.4\times 10^{-8} \,.
\end{eqnarray}
Notice that these bounds are quoted at the $90 \%$ CL so they will be rescaled to $1 \sigma$ to build the corresponding contribution to the $\chi^2$ function.

%%%%%%

\section{Results}
\label{sec:res}

%%%%%%%%%%%%%%%%%%%%%%%%%
\subsection{Constraints from the global fit}
%%%%%%%%%%%%%%%%%%%%%%%%%

With the 13 observables discussed in Section~\ref{sec:obs} we build a $\chi^2$ function depending on the 9 parameters listed in Table~\ref{tab:params}. Given the large dimensionality of the parameter space, we make use of Markov chain Monte Carlo (MCMC) techniques for efficient parameter exploration. In particular, we implement importance sampling based on the Likelihood obtained from the observables through a Metropolis-Hastings algorithm. The range in which the 9 free parameters are varied is also summarized in Table~\ref{tab:params}. We have run simultaneously 5 different chains through the MCMC algorithm and have verified that good convergence (better than $R-1 < 0.035$~\cite{Gelman:1992zz}) for all parameters has been achieved. The results of the runs thus provide a good sample of the $\chi^2$ values in the preferred regions of the parameter space and have been used to marginalize over different subsets of the model parameters. In this way, we will present 2D and 1D frequentist contours on the more phenomenologically relevant parameters of the model. The post-processing of the chains to derive the allowed confidence regions has been performed with the MonteCUBES~\cite{Blennow:2009pk} user interface.

\begin{figure}
\centering
\includegraphics[width=0.46\textwidth]{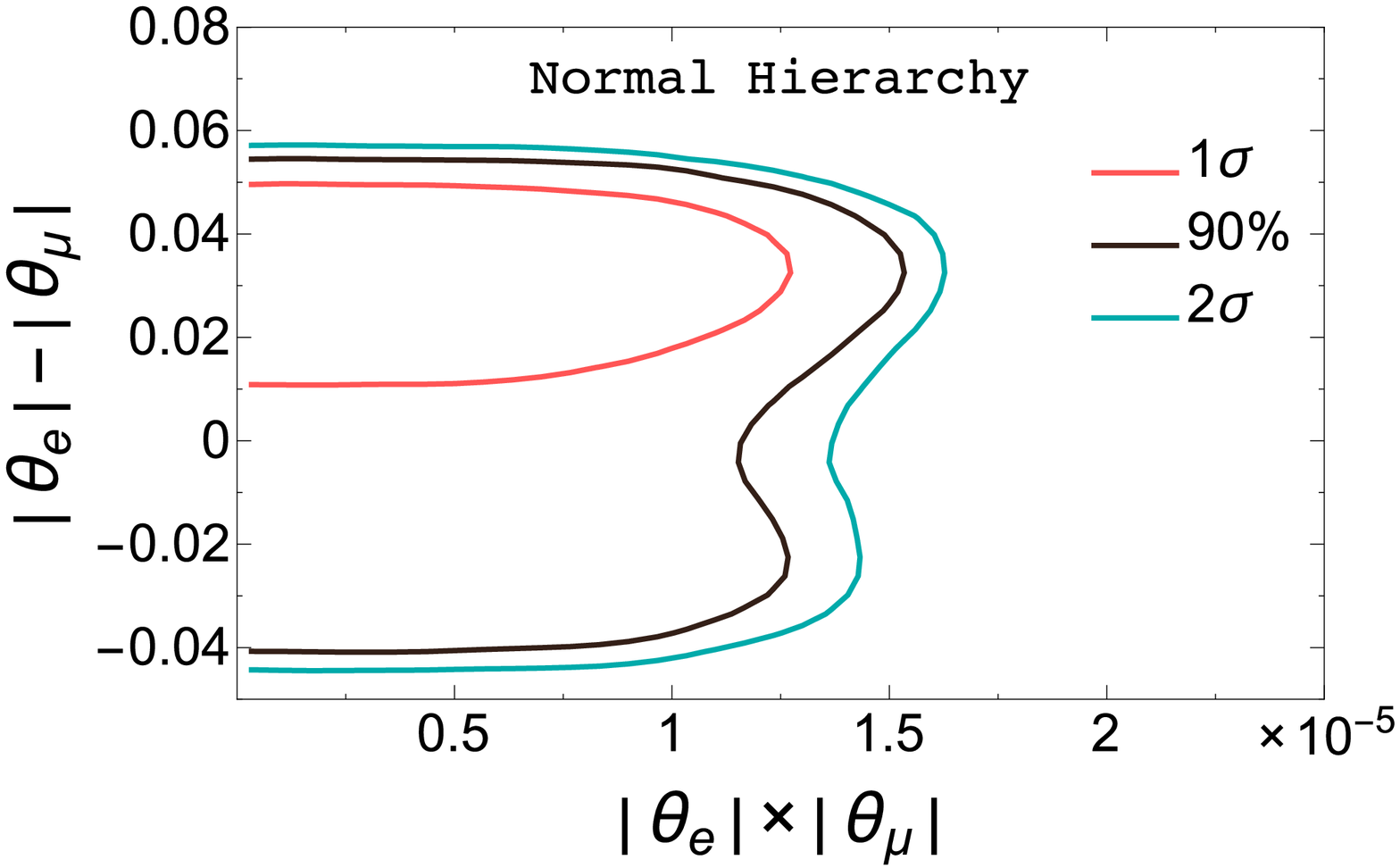}
\includegraphics[width=0.46\textwidth]{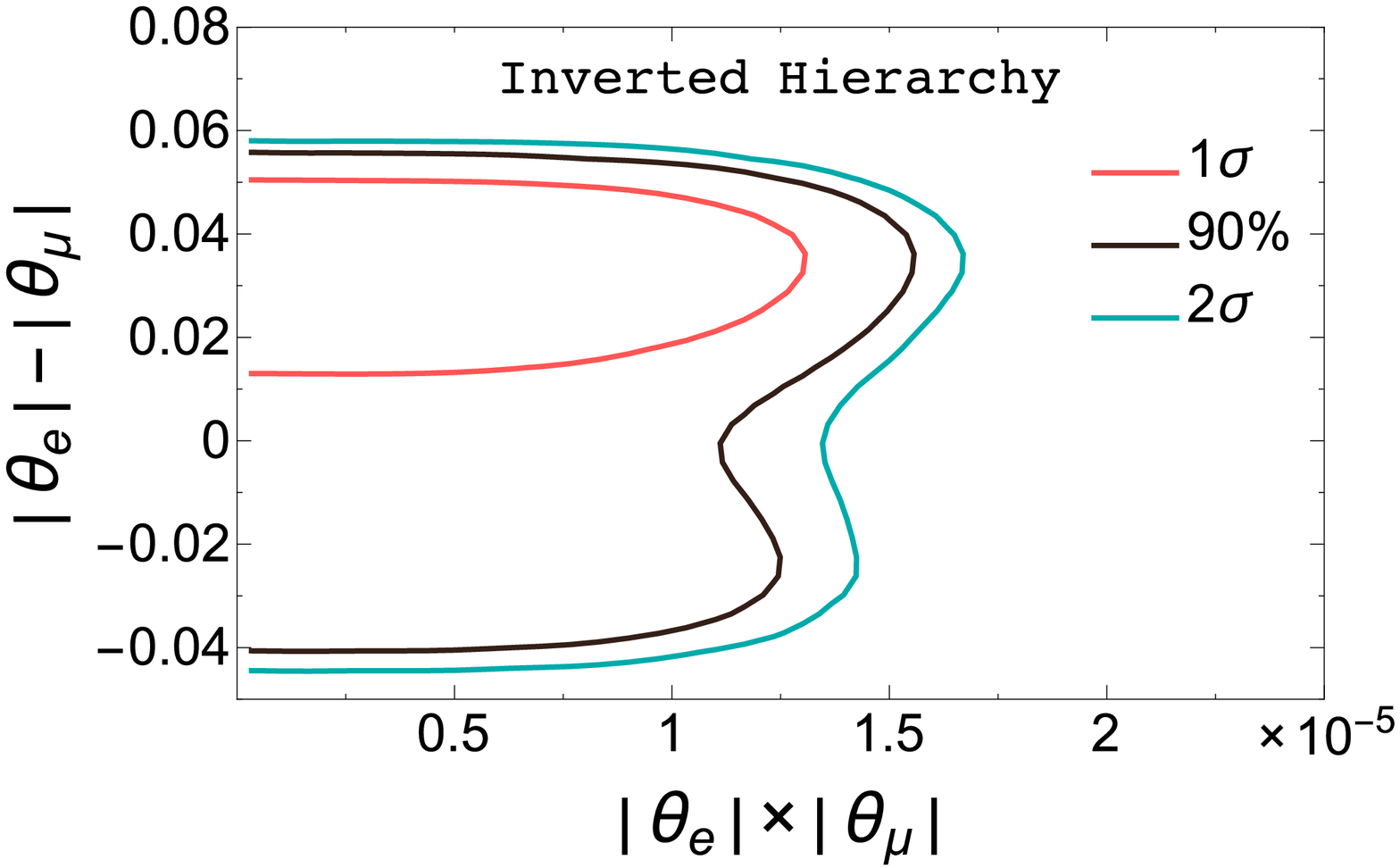}
\includegraphics[width=0.45\textwidth]{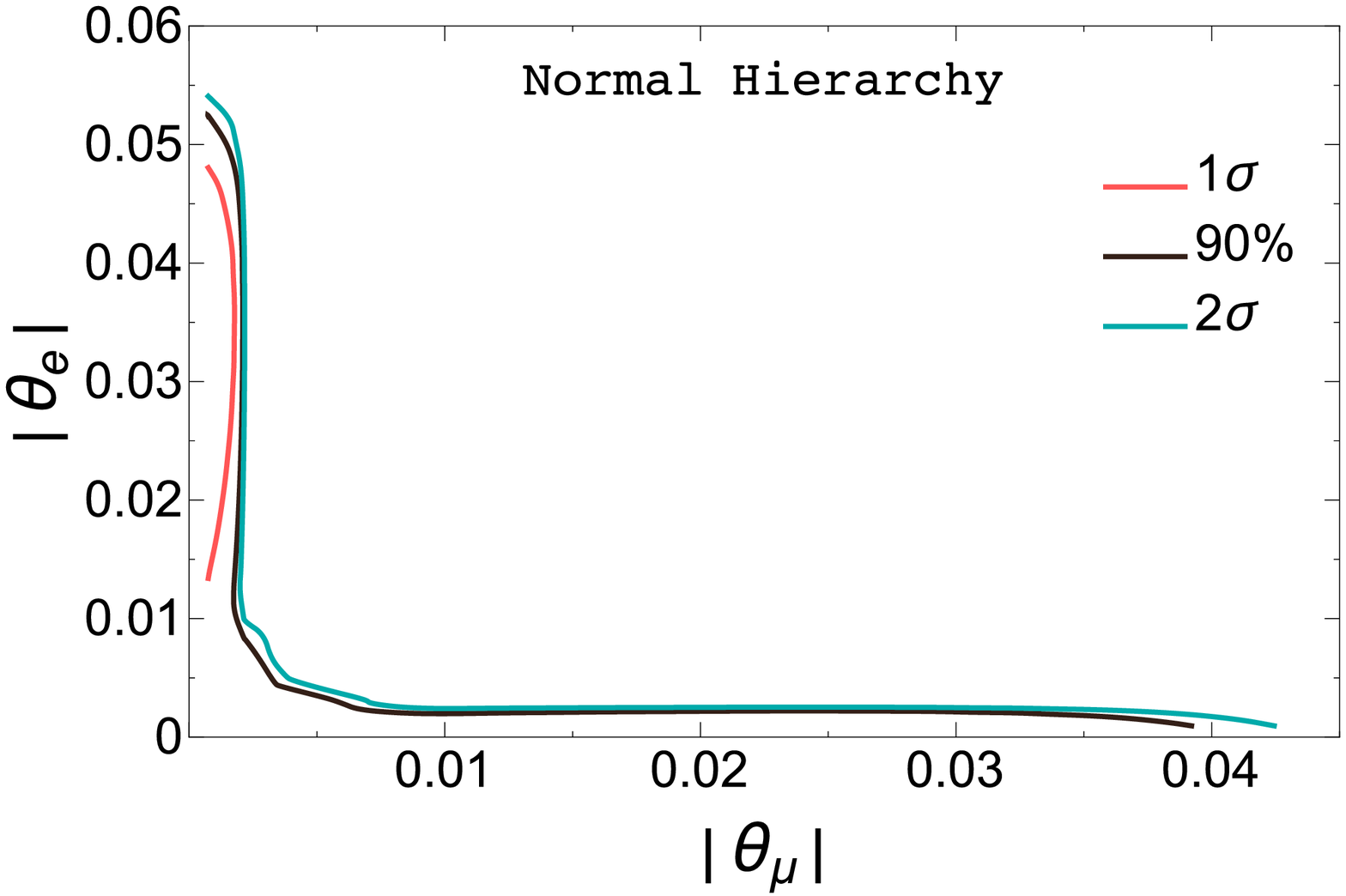}
\includegraphics[width=0.45\textwidth]{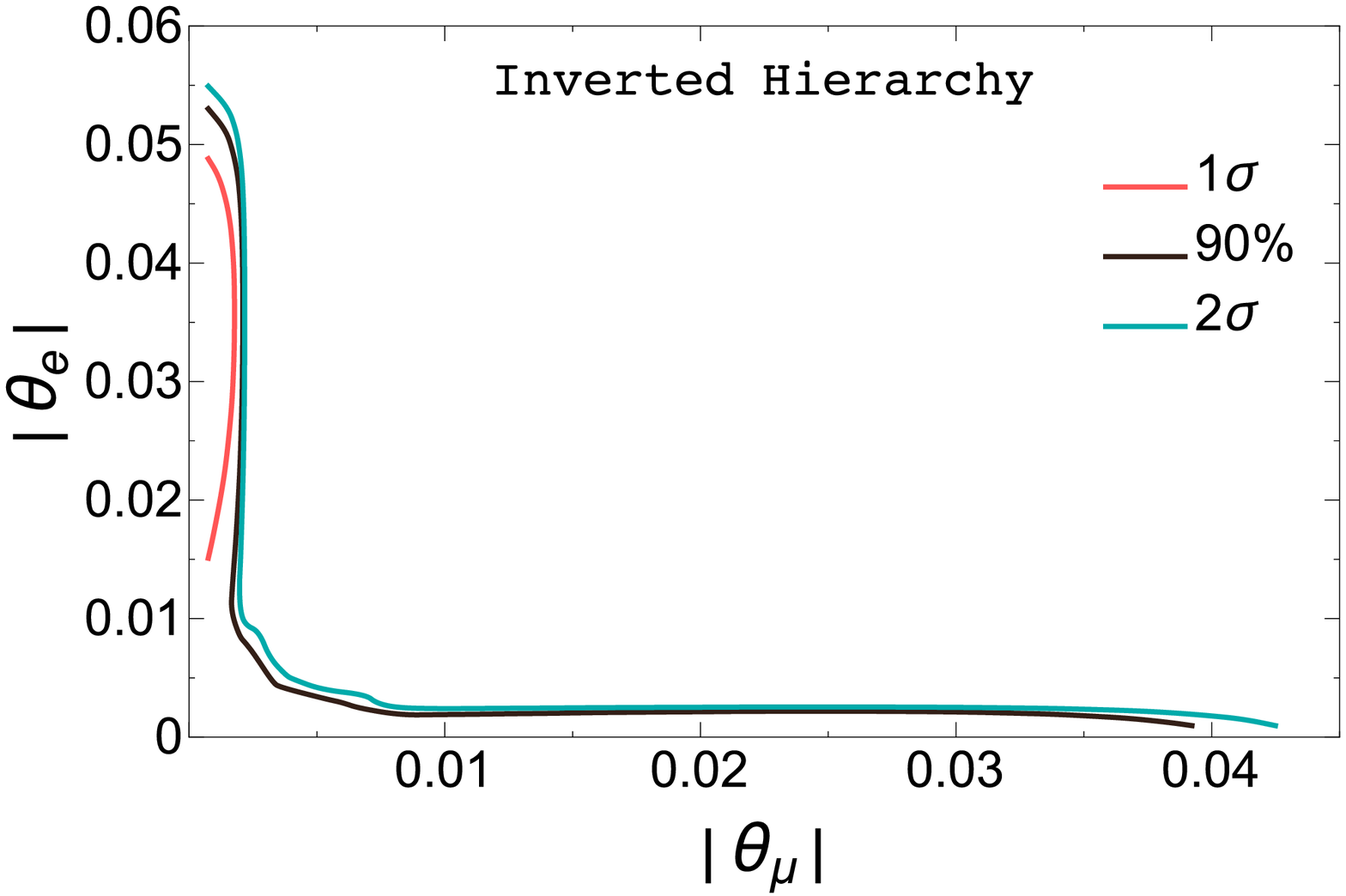}
\includegraphics[width=0.45\textwidth]{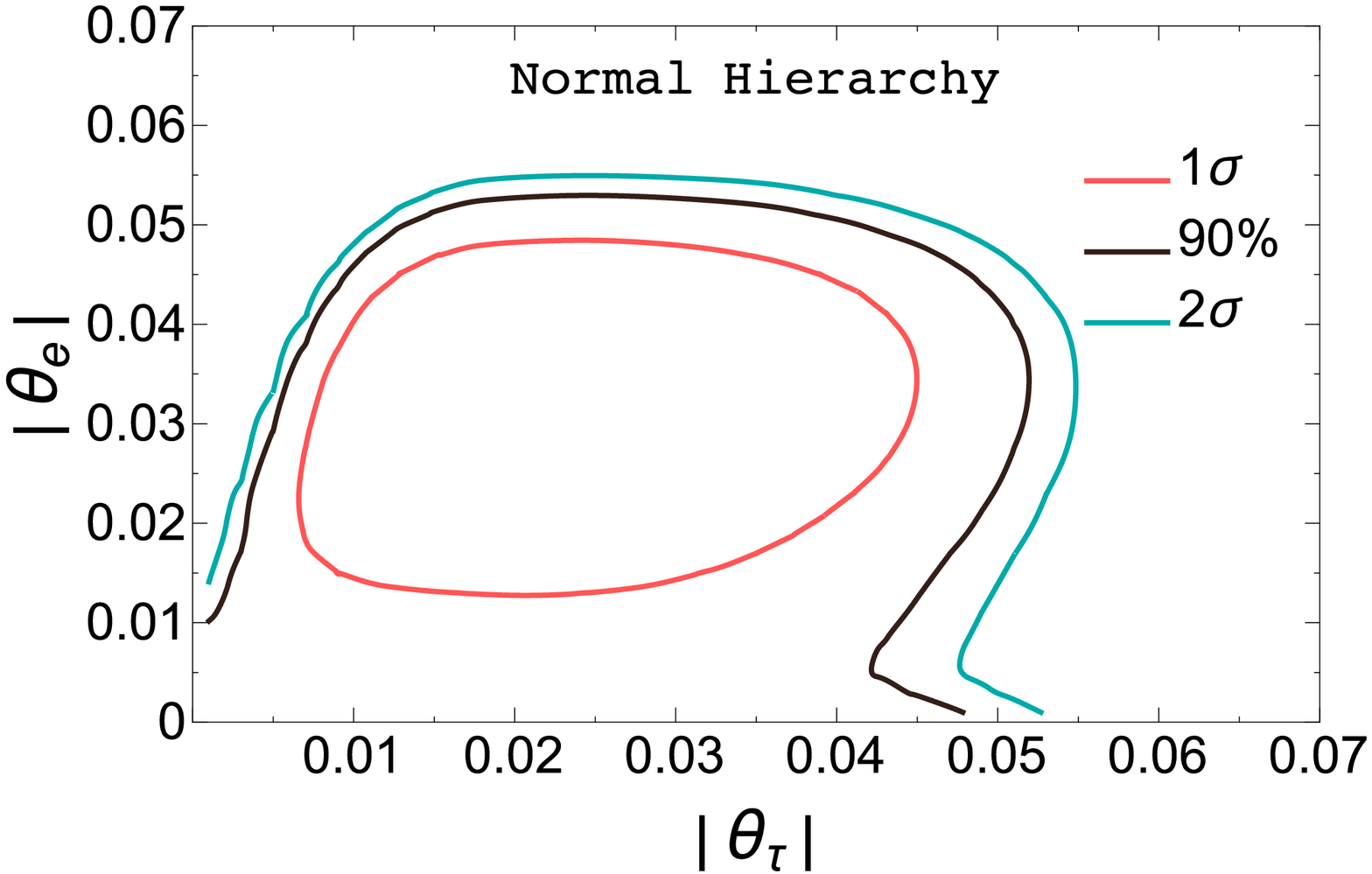}
\includegraphics[width=0.45\textwidth]{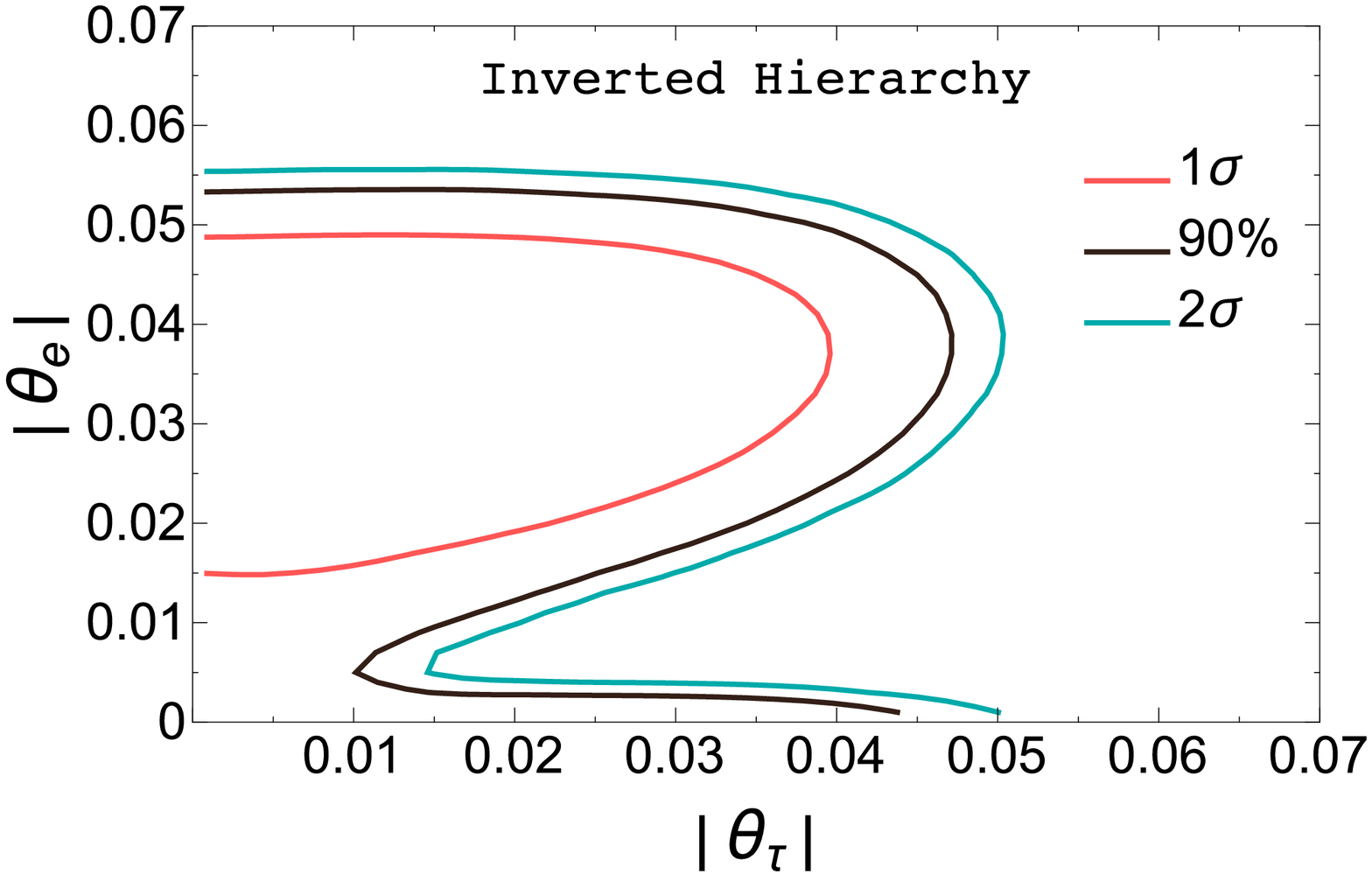}
\caption{Contours for $\theta_{e}$, $\theta_{\mu}$ and $\theta_{\tau}$ at $1\sigma$ (red), $90\%$ CL (black) and $2\sigma$ (blue). The left panels are obtained for normal hierarchy and the right for inverted.}
\label{fig:contours}
\end{figure}

In Fig.~\ref{fig:contours} we show the results of our MCMC scan for the 2 degrees of freedom constraints of different combinations of the heavy-active mixings $\theta_{\alpha}$ defined in Eq.~(\ref{eq:theta}). The contours correspond to the $1 \sigma$, $90 \%$ and $2 \sigma$ frequentist confidence regions. The upper panels show the bounds in the two combinations we choose to more directly sample (see Table~\ref{tab:params}): $|\theta_e| \times |\theta_\mu|$ and $|\theta_e| - |\theta_\mu|$. The rationale behind this is apparent upon inspection of Fig.~\ref{fig:contours}. Indeed, the constraints on the product are more than one order of magnitude smaller than those derived from the difference of the couplings $\sqrt{|\theta_e| \times |\theta_\mu|} \ll ||\theta_e| - |\theta_\mu||$, leading to a very pronounced hyperbolic degeneracy in the panels of the middle row, which contain the same information directly depicted as a function of $\theta_e$ and $\theta_\mu$. Thus, this particular choice of sampling parameters allowed to scan the hyperbolic degeneracy much more efficiently and speed the convergence of the MCMC. This very strong constraint in $|\theta_e| \times |\theta_\mu|$ stems from the strong bound on $\mu \to e \gamma$ from MEG that, from Eq.~(\ref{eq:raredecay}), sets a very stringent limit on $|\theta_\mu \theta_e^*|$. 

Finally, the lower panels of Fig.~\ref{fig:contours} contain the constraints derived for the mixing with the $\tau$ flavour $\theta_\tau$. Notice that $Y_\tau$, and hence $\theta_\tau$, was not a free parameter of the fit but was rather obtained from the other two Yukawas and the light neutrino masses and mixings from Eq.~(\ref{eq:Yt}). This is the source of the observed correlation between the values of $\theta_e$ and $\theta_\tau$. Notice also that, since the particular pattern of light neutrino masses plays an important role in Eq.~(\ref{eq:Yt}), the left (normal hierarchy) and right (inverted hierarchy) panels of Fig.~\ref{fig:contours} display different correlations. 

\begin{figure}
\centering
\includegraphics[width=0.465\textwidth]{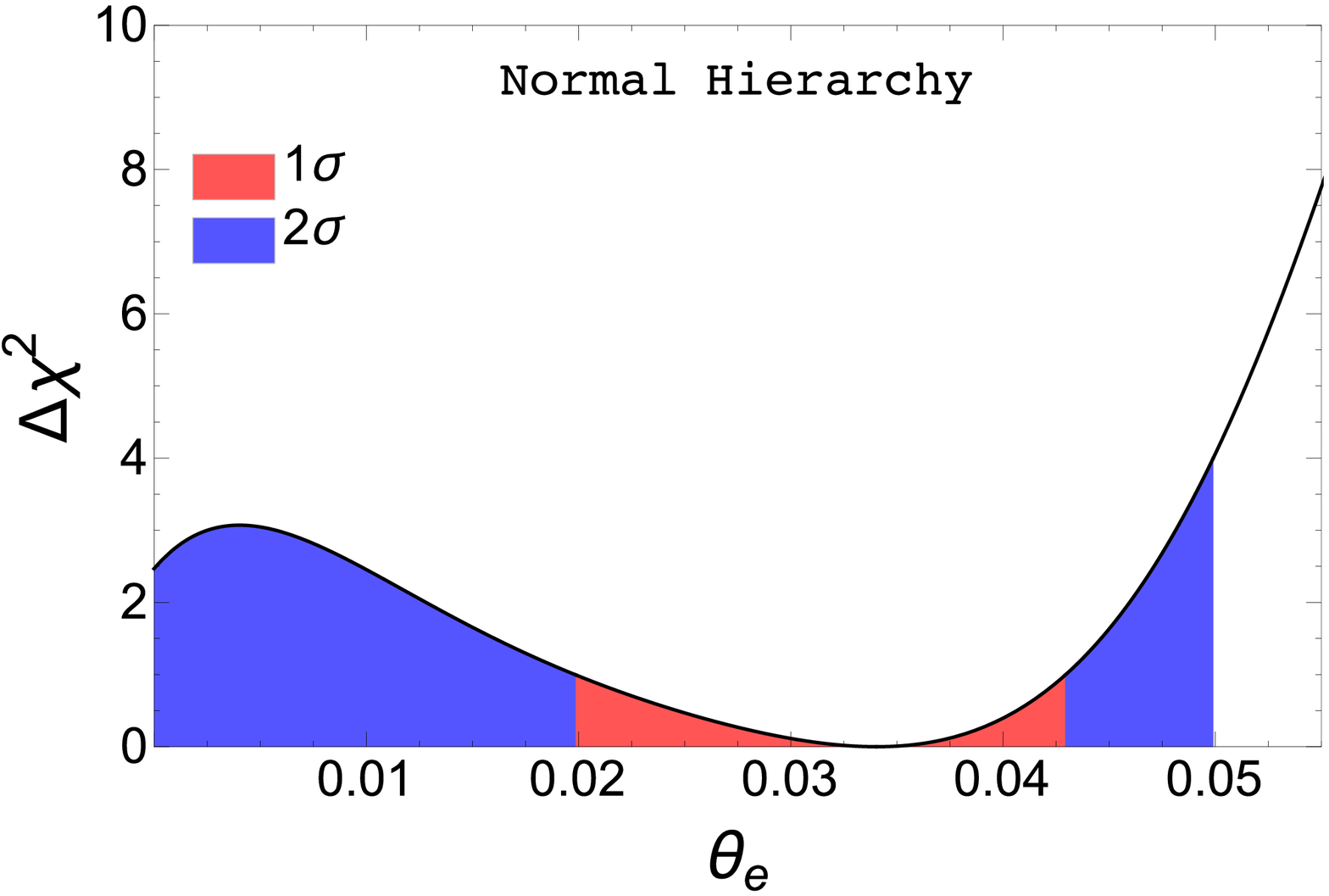}
\includegraphics[width=0.465\textwidth]{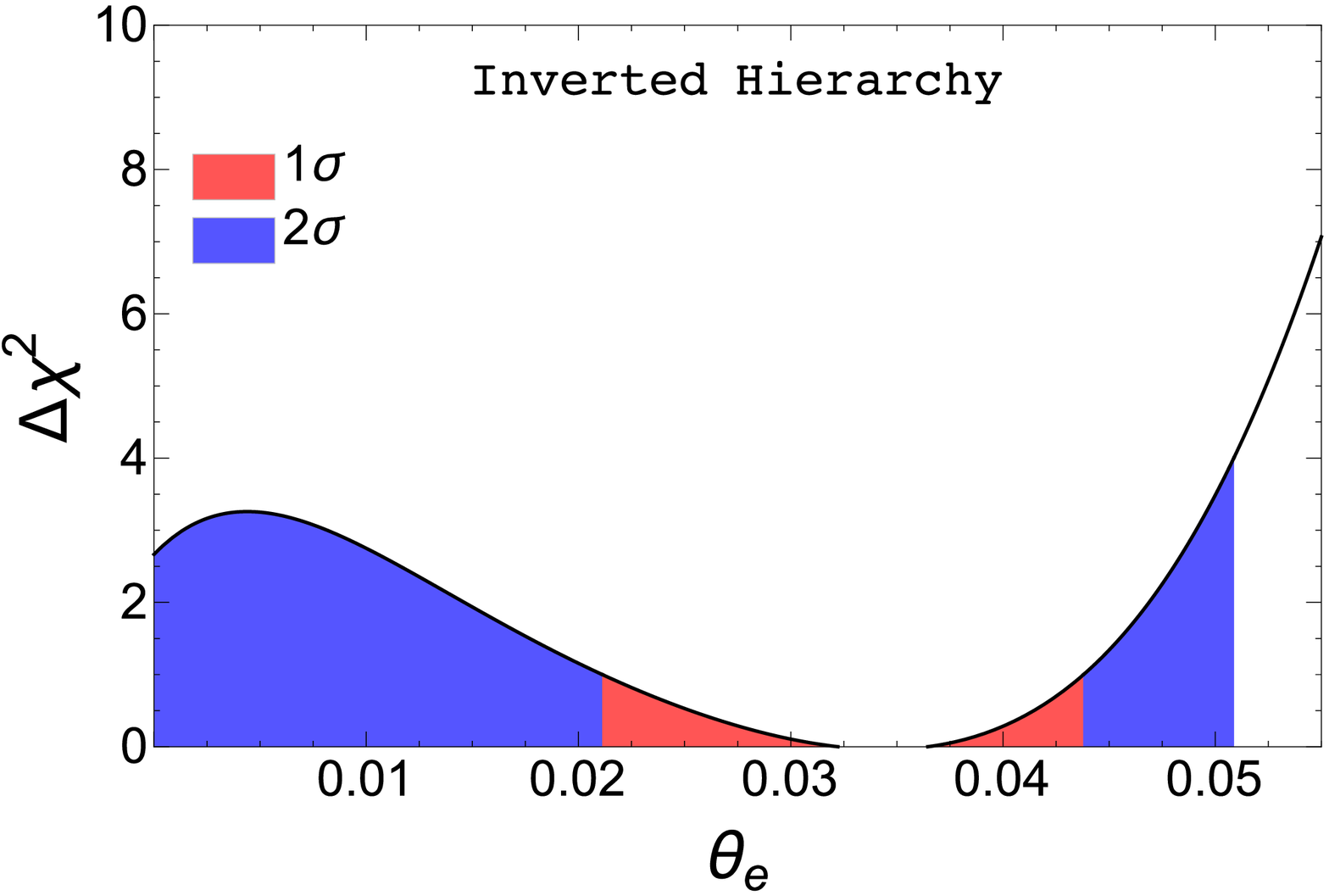}
\includegraphics[width=0.47\textwidth]{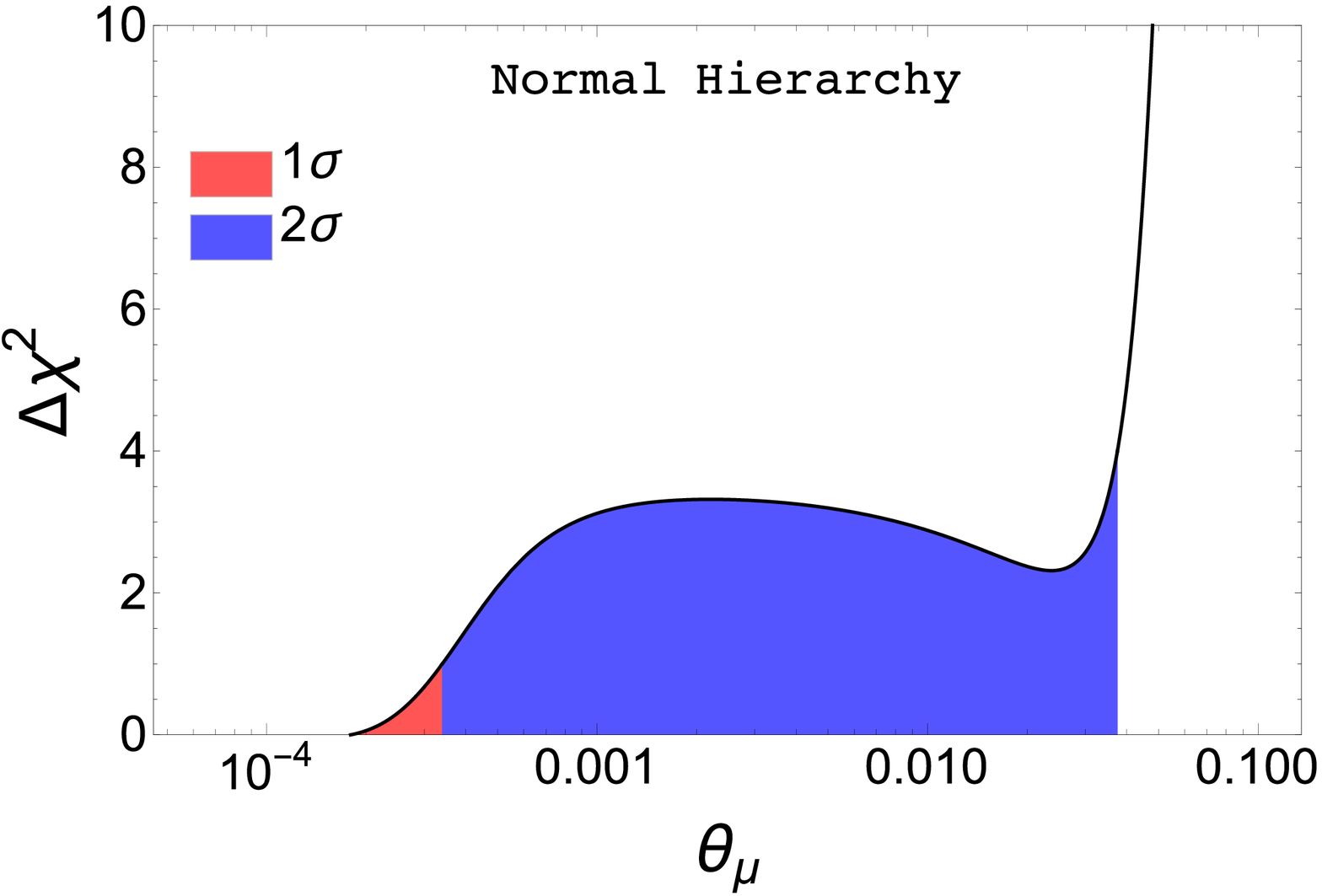}
\includegraphics[width=0.47\textwidth]{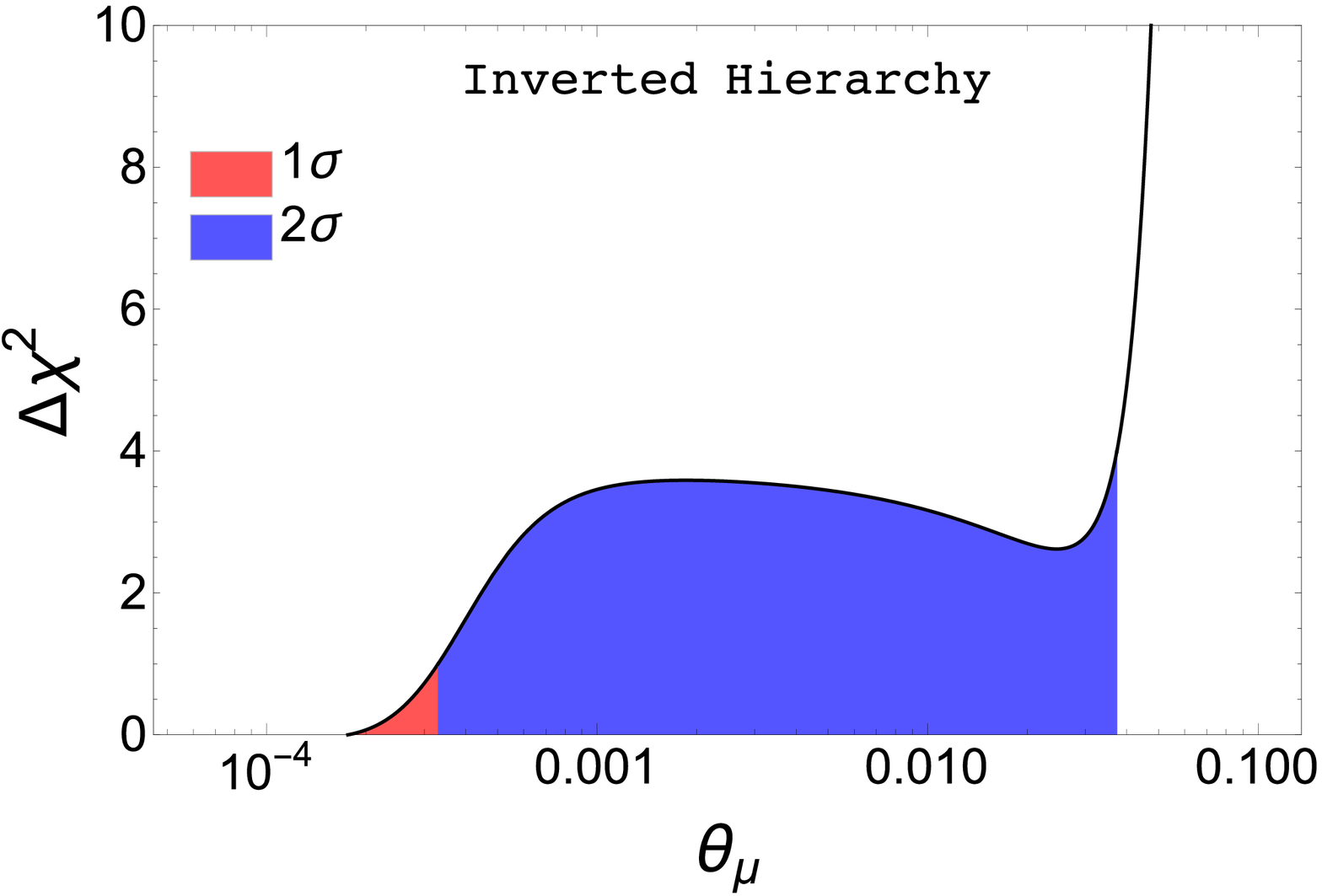}
\includegraphics[width=0.465\textwidth]{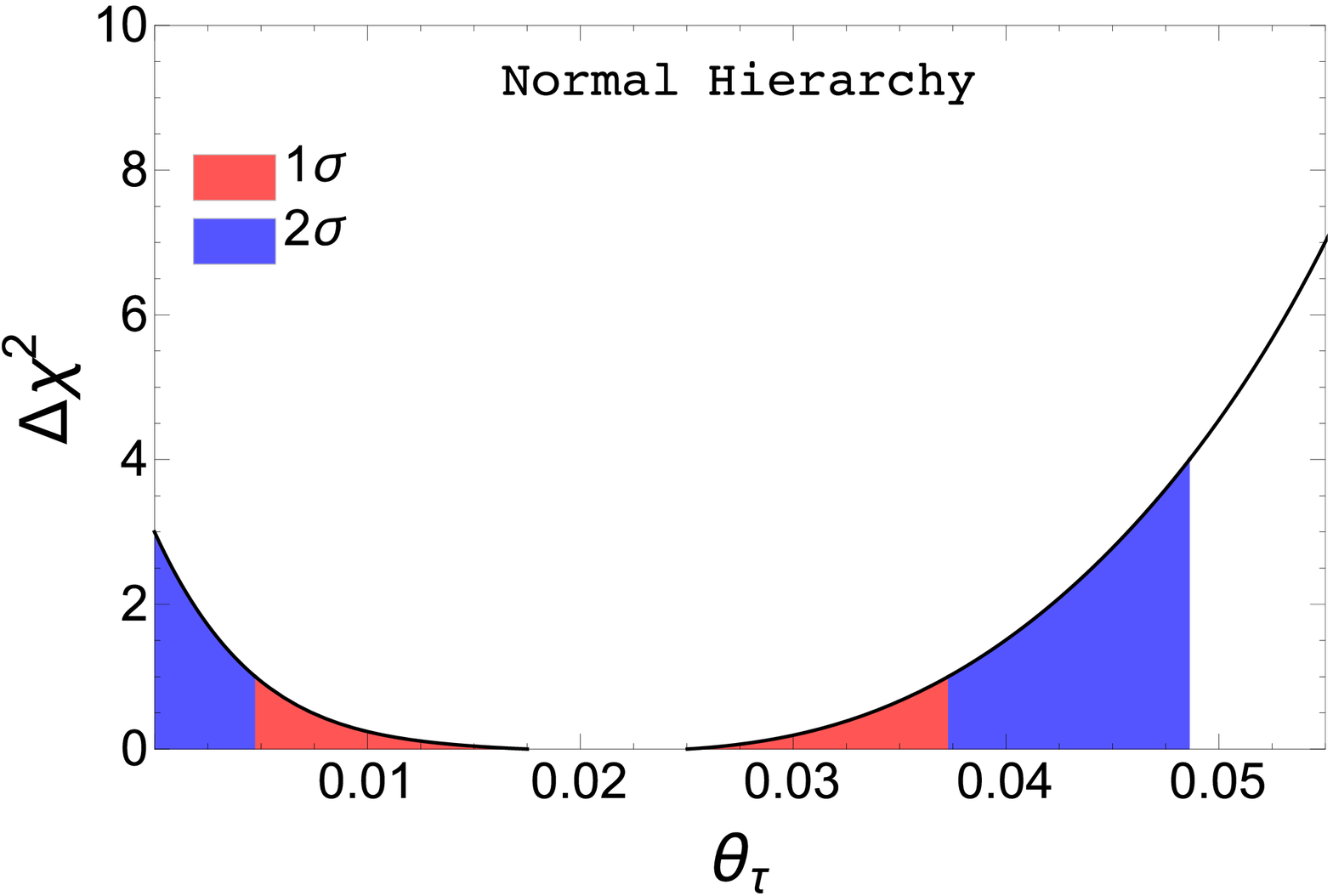}
\includegraphics[width=0.465\textwidth]{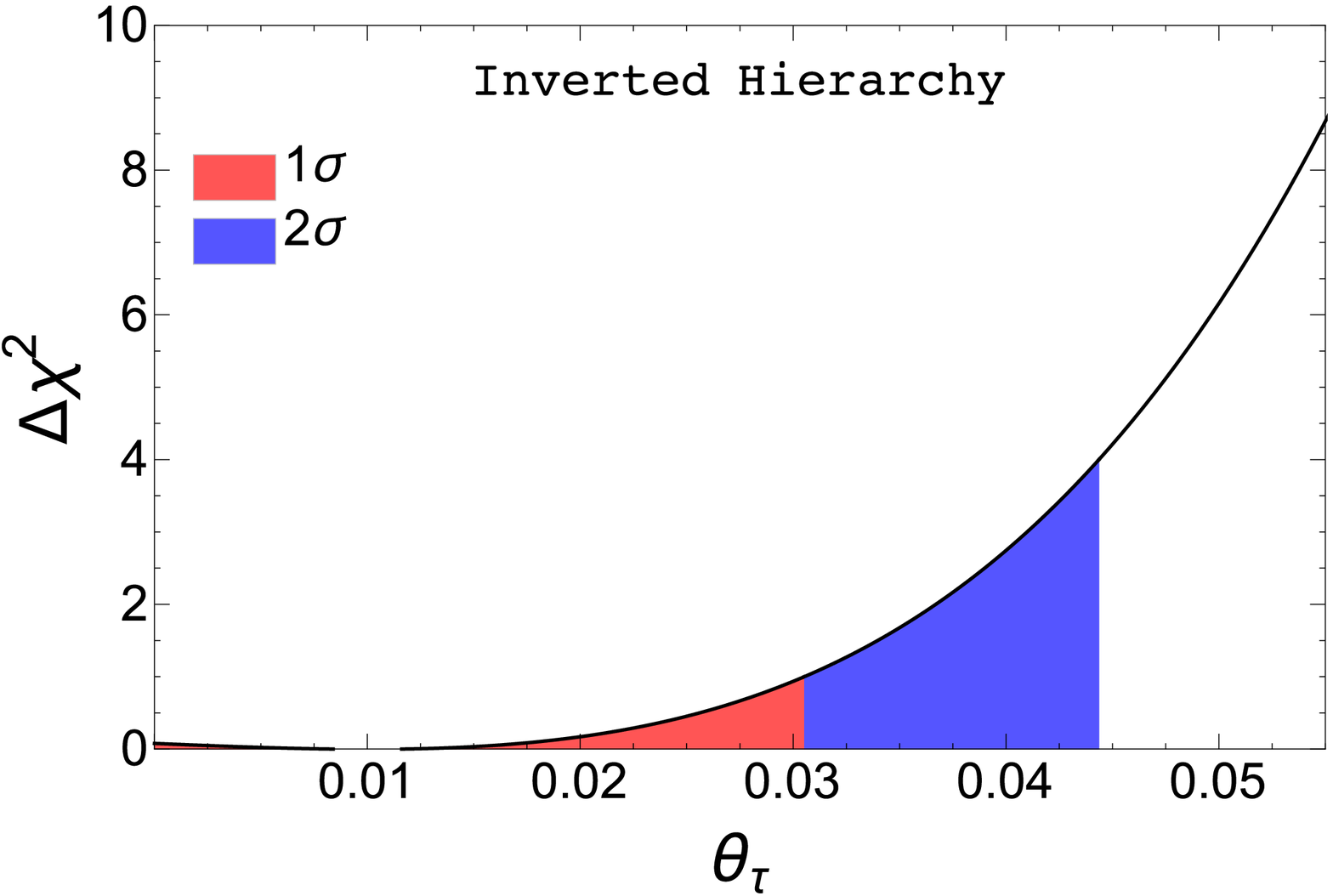}
\caption{$\Delta \chi^{2}$ (marginalized over all other parameters) for $\theta_{e}$, $\theta_{\mu}$ and $\theta_{\tau}$. Left panels show results for a normal hierarchy and right panels for inverted hierarchy.}
\label{fig:1D}
\end{figure}

In Fig.~\ref{fig:1D} we show the individual constraints that can be derived on $\theta_e$, $\theta_\mu$, and $\theta_\tau$ (from top to bottom) for a normal (left) and an inverted (right) hierarchy after marginalizing over all other parameters. We generally find a slight improvement of the fit to the observables considered when some amount of mixing is present. In particular, we find that non-zero mixing with the electron is preferred at around the $90 \%$~CL by our dataset. Mixing with the tau flavour is also favoured for normal hierarchy due the correlations implied by Eq.~(\ref{eq:Yt}). At the $1 \sigma$ level, mixing with the $\mu$ flavour is significantly constrained due to the preference of some universality bounds (from $\pi$ and $\tau$ decays) for a slightly reduced coupling to the electron with respect to the muon. Thus, since universality constraints are corrected by $1-|\theta_\alpha|$ for each flavour, a non-zero $\theta_e$ is preferred in the fit while $\theta_\mu$ is kept at small values to satisfy the constraint from $\mu \to e \gamma$. Beyond the $1 \sigma$ level, the mixing with the electron is allowed to become small and thus the constraint on $\mu$ mixing at $2 \sigma$ is much weaker than naively expected from the $1 \sigma$ region. The limits of the 1 and $2 \sigma$ regions for the three mixing parameters are summarized in Table~\ref{tab:constraints}. 

\begin{table}[htb!]
\centering
\begin{tabular}{|c|c|c|c|c|c|c|}
\hline
    & \multicolumn{2}{|c|}{$\theta_e$} & \multicolumn{2}{|c|}{$\theta_\mu$} & \multicolumn{2}{|c|}{$\theta_\tau$} \\
\hline
    & $1 \sigma$ & $2 \sigma$ & $1 \sigma$ & $2 \sigma$ & $1 \sigma$ & $2 \sigma$ 
 \\
\hline
NH & $0.034^{+0.009}_{-0.014}$ & $<0.050$ & $<3.2 \cdot 10^{-4}$ & $<0.037$ &$0.018^{+0.019}_{-0.013}$ & $<0.049$   \\
\hline
IH & $0.035^{+0.009}_{-0.014}$ & $<0.051$ & $<3.3 \cdot 10^{-4}$ & $<0.037$ & $<0.031$ & $<0.044$   \\
\hline
\end{tabular}
\caption{Constraints on $\theta_e$, $\theta_\mu$, and $\theta_\tau$ for normal and inverted hierarchy.}\label{tab:constraints}
\end{table}

\begin{figure}
\centering
\includegraphics[scale=0.55,angle=270]{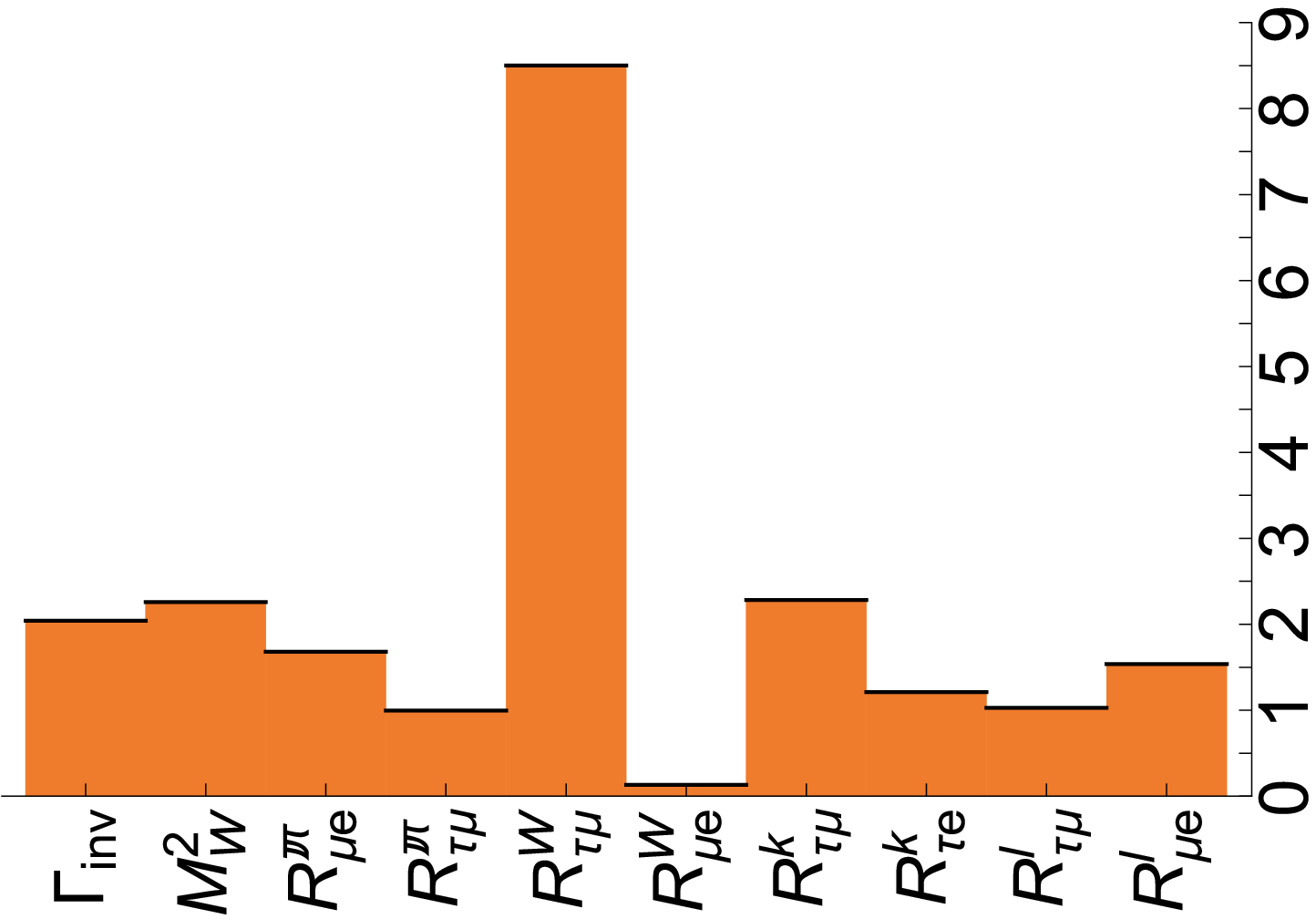}
\includegraphics[scale=0.55,angle=270]{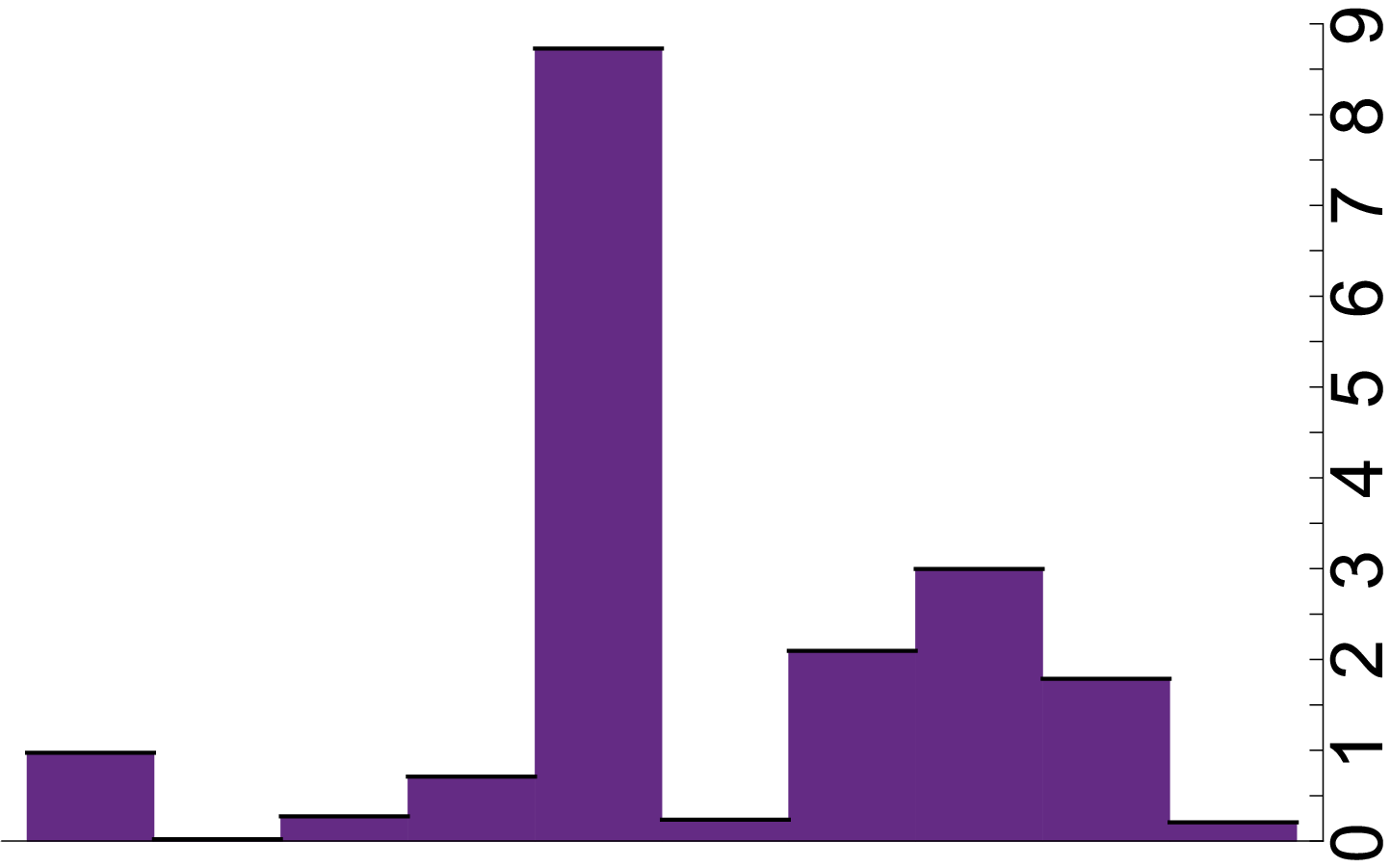}
\includegraphics[scale=0.55,angle=270]{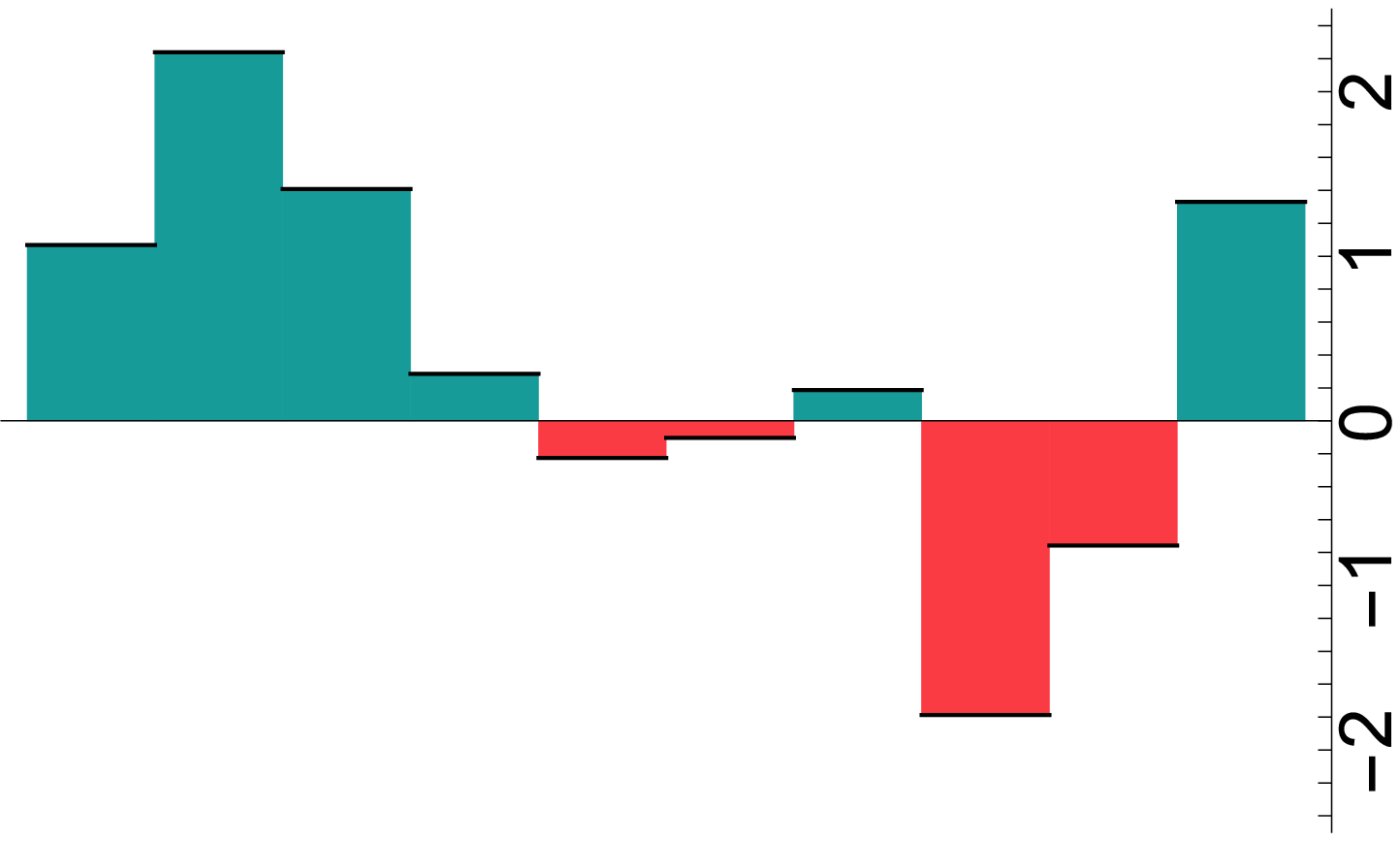}
${\chi^2(SM)}$ \hspace{3.5cm} $\chi^2(BF)$ \hspace{3.5cm} $\Delta\chi^{2}(SM)$
\caption{Contributions from the different observables to the $\chi^{2}$. Left plot shows the SM values. Middle plot shows the contributions from three right-handed neutrinos in the best-fit point. Right plot shows $\Delta\chi^2_i\equiv\chi^{2}_{i}(SM)-\chi^{2}_{i}(BF)$ for every observable $i$.}
\label{fig:chi2}
\end{figure}

In Fig.~\ref{fig:chi2} we show a comparison of the breakdown of the contributions of the different observables to the total $\chi^2$ for the SM (left panel) and 
our best fit (middle panel) as well as the difference of the two (right panel). It can be seen that some of the existing tension of the SM with the invisible width 
of the $Z$ can be alleviated by the presence of heavy neutrino mixings and also the agreement between the kinematic determination of $M_W$ and its SM value 
from $G_F$, $\alpha$ and $M_Z$ is improved. As already discussed, the universality constraints from $\pi$ and $\tau$ decays are also in better agreement when 
some mixing with the electron is present. On the other hand, universality tests from kaon decays rather point in the opposite direction. Thus, at the end, the 
preference for non-vanishing heavy-active mixing is mild and the final improvement of the $\chi^2$ with respect to the SM value is 3.7, not quite reaching the $2 \sigma$ level. Notice that, even if the number of free parameters in the fit is rather high, the observables actually depend 
on the combinations $|\theta_e|$, $|\theta_\mu|$ and $|\theta_\tau|$ only (and $\Lambda$ when loop corrections are relevant). Thus, the reduction by 3.7 of the 
$\chi^2$ should be attributed to the introduction of 3 (or 4) new parameters rather than 9.

Regarding the importance of the loop effects considered, we have performed a second set of MCMC runs where all loop corrections have been removed. The results of these simulations are essentially identical to the ones stemming from the full computation. By adding to the chain output also the value that the $T$ parameter took in the simulations, we find that its preferred values are $\sim 10^{-7}-10^{-6}$, negligible with respect to the best fit values of the tree level contributions. In order to understand this apparent lack of relevance of the loop corrections and the $T$ parameter in particular, in direct contrast to the results presented in~\cite{Akhmedov:2013hec}, we will now analyze in further detail the regions of the parameter space in which $T$ could be relevant and the necessary conditions for the cancellation with the tree level contributions to take place.

%%%%%%%%%%%%%%%%%%%%%%%%
\subsection{The $T$ parameter}
%%%%%%%%%%%%%%%%%%%%%%%%

The leading contributions (not suppressed by the light neutrino or charged lepton masses) to the $T$ parameter are given by~\cite{Akhmedov:2013hec}:
\begin{equation}
\alpha T = \frac{\alpha}{8 \pi s_\mathrm{W}^2 M^2_W}\left(  \sum_{\alpha,\beta,i,j}\left(U^*_{\alpha i} U_{\alpha j} U_{\beta i} U^*_{\beta j} f_{ij} +  U^*_{\alpha i} U_{\alpha j} U^*_{\beta i} U_{\beta j} g_{ij}\right) \right),
\label{eq:T}
\end{equation}
where
\begin{equation}
f_{ij} = \frac{M_i^2 M_j^2}{M_i^2-M_j^2}\ln{\frac{M_i}{M_j}}
\qquad
\mathrm{and}
\qquad
g_{ij} = \frac{2 M_i M_j^3}{M_i^2-M_j^2}\ln{\frac{M_i}{M_j}} ,
\end{equation}
and where $M_i$ are the neutrino mass eigenvalues. In~\cite{Loinaz:2002ep,Loinaz:2004qc} it was shown that several of the most constraining observables, notably the $Z$ decay to charged leptons and $\sin^2 \theta^{\rm eff}_w$~\cite{ALEPH:2002aa}, depended on the combination:
\begin{equation}
(NN^\dagger)_{ee} (NN^\dagger)_{\mu \mu} - 2 \alpha T \simeq 1 - |\theta_e|^2 - |\theta_\mu|^2 - 2 \alpha T .
\label{eq:cancellation}
\end{equation}
Since from Table~\ref{tab:constraints} $|\theta_e|^2 + |\theta_\mu|^2 \sim 10^{-3}$, $2 \alpha T$ must be of similar order so as to be competitive with the tree contribution. From Eq.~(\ref{eq:T})
\begin{equation}
2 \alpha T \simeq \frac{\alpha \Lambda^2 |\theta_\alpha|^4}{16 \pi s_\mathrm{W}^2 M^2_W},
\end{equation}
where $\Lambda$ is the mass scale of the heavy neutrinos and $\theta_\alpha/\sqrt{2}$ their mixing with the flavour states from Eq.~(\ref{eq:theta}). Thus, in order for $2 \alpha T \sim |\theta_\alpha|^2$ it is necessary that $\Lambda \sim 10-100$~TeV. And, since $|\theta_\alpha|^2 \sim |Y_\alpha|^2 v_\text{EW}^2/2 \Lambda^2 \sim 10^{-3}$, then $|Y_\alpha| \sim 1-10$, on the very limit of perturbativity but, a priori, an interesting possibility.

Furthermore, notice that the second term in Eq.~(\ref{eq:T}) has the typical structure in the elements of the mixing matrix $U$ of $L$-violating processes, such as, for example, neutrinoless double $\beta$ decay. Indeed, this term stems from the correction to the $Z$ propagator with two neutrinos running in the loop and a Majorana mass insertion and it is easy to see that it vanishes in the limit of exactly conserved Lepton number, taking all $\epsilon_i$ and $\mu_j$ to zero. Thus, if $B-L$ is approximately conserved, the first term in Eq.~(\ref{eq:T}) dominates the contribution to $T$. However, it can be shown that the matrix $f_{ij}$ is positive semi-definite for three extra heavy neutrinos or less\footnote{Preliminary explorations indicate that this argument can be generalized to more extra heavy neutrinos.} and can then be diagonalized as $f_{ij} = \sum_k V_{ik} \lambda_k V^*_{jk}$, where $V$ is a Unitary matrix and $\lambda_k \geq 0$. Thus, if $B-L$ is approximately conserved:
\begin{equation}
\alpha T  \sim  \frac{\alpha}{8 \pi s_\mathrm{W}^2 M^2_W}  \sum_{\alpha, \beta, i} \left| \sum_{k} U^*_{\alpha i} U_{\beta i}  V_{ik} \sqrt{\lambda_k} \right|^2 \geq 0 .
\end{equation}
But from Eq.~(\ref{eq:cancellation}) $T < 0$ is mandatory so as to have the cancellation between $T$ and $|\theta_\alpha|^2$ discussed in~\cite{Akhmedov:2013hec}. Thus, significant violations of $B-L$ are necessary so that the second term in Eq.~(\ref{eq:T}), which is allowed to be negative, can dominate over the first. 

\begin{figure}
\centering
\includegraphics[scale=0.6]{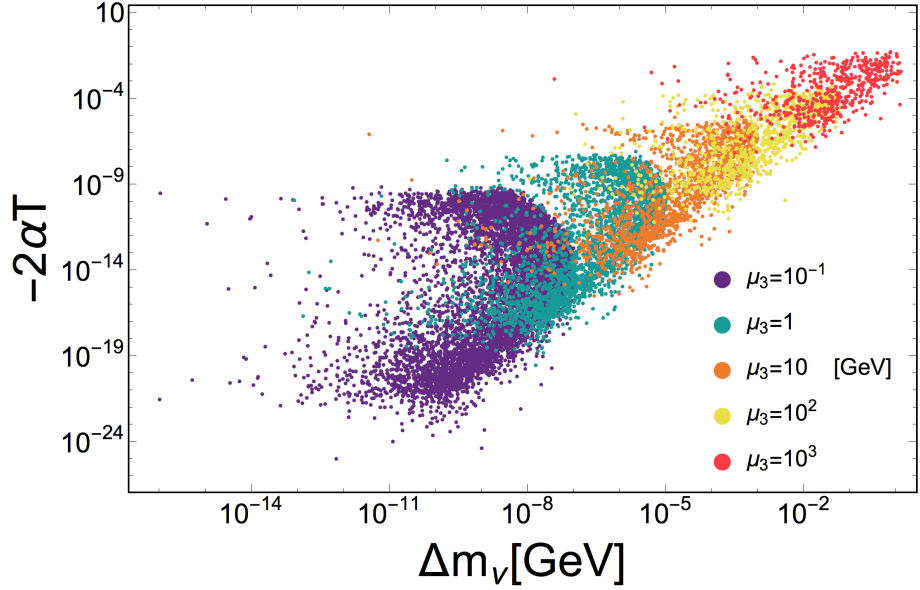}
\caption{$T$ parameter versus 1-loop correction to $m_\nu$ for different values of the $L$-violating parameters $\mu_1$ and $\mu_3$.}
\label{fig:T}
\end{figure}

Notice that, for arbitrary values of the $B-L$-violating parameters $\epsilon_i$ and $\mu_j$, Eq.~(\ref{eq:texture}) is a completely general parametrization of 
a type-I Seesaw mechanism with three extra right-handed neutrinos. But, given Eq.~(\ref{eq:lightmass0}), only $\mu_1$ and $\mu_3$ are allowed to be sizable 
given the present constraints on the light neutrino masses and mixings. If $|\mu_1| \gg \Lambda,\Lambda',\mu_3$ a negative $T$ can indeed be obtained:
\begin{equation}
T\simeq \frac{v_\text{EW}^4}{32\pi s_\mathrm{W}^2 M_W^2 \mu_1^2}\left(\displaystyle\sum_\alpha\left|Y_\alpha\right|^2\right)^2\left(3-4\log\left(\frac{\mu_1}{\Lambda}\right)\right) .
\end{equation}
If both $\mu_1$ and $\mu_3$ are simultaneously included and dominate over the $L$-conserving $\Lambda$ and $\Lambda'$ then $T$ is given by: 
\begin{equation}
T\simeq \frac{v_\text{EW}^4}{64\pi s_\mathrm{W}^2 M_W^2}\left(\displaystyle\sum_\alpha\left|Y_\alpha\right|^2\right)^2 \frac{6 \mu \mu_1 - \left( 3 \mu_1^2 + \mu^2 \right) \log\left(\frac{\mu + \mu_1}{\mu - \mu_1}\right)}{\mu^3 \mu_1}  ,
\label{eq:TLV}
\end{equation}
where $\mu = \sqrt{\mu_1^2 + 4 \mu_3^2}$. In this limit, negative values of $T$ are also easily accessible. However, the price to pay is high, the approximate $B-L$ symmetry protecting the Weinberg operator despite the Yukawas at the very border of perturbativity and the low Seesaw scale, has been strongly broken by $\mu_1$ and $\mu_3$. While this does not induce any dangerous corrections to neutrino masses at tree level, and hence when working with the Casas-Ibarra parametrization as in Ref.~\cite{Akhmedov:2013hec} the correct masses and mixings seem to be recovered, loop corrections need to also be taken into account since no protecting symmetry can now suppress them. Indeed, the loop contributions mediated by $\mu_1$ and $\mu_3$ to the light neutrino masses are found to be~\cite{Pilaftsis:1991ug,Grimus:2002nk,AristizabalSierra:2011mn,LopezPavon:2012zg,Dev:2012sg}:
\begin{equation}
\Delta m_{\nu_{\alpha\beta}}
= \frac{Y_\alpha Y_\beta}{32\pi^2\mu}\left(3M_Z^2f(M_Z)+M_h^2f(M_h)\right) \, ,
\end{equation}
with: 
\begin{equation}
f(M)=\frac{(\mu + \mu_1)^2\log{\left(\frac{\mu + \mu_1}{2M}\right)}}{\left( \mu + \mu_1 \right)^2-4M^2}-\frac{(\mu - \mu_1)^2\log{\left(\frac{\mu - \mu_1}{2M}\right)}}{\left(\mu - \mu_1 \right)^2-4M^2} \, .
\label{eq:f}
\end{equation}
These corrections can indeed be sizable and in Fig.~\ref{fig:T} we show the values that the loop contribution to the light neutrino masses take in order to recover a given value for $-2 \alpha T$ for different values of $\mu_1$ and $\mu_3$. From inspection of Eq.~(\ref{eq:f}), the limit of vanishing $\mu_1$ would render $f(M) = 0$, keeping under control the loop corrections to neutrino masses\footnote{In this limit with $\mu_3 \gg \Lambda,\Lambda'$, $L$-symmetry is recovered with two degenerate neutrinos with mass $\mu_3$ that form a Dirac pair. Hence, the symmetry ensures the stability of $\nu$ masses at loop level but conversely drives $T$ to positive values.}. However, from Eq.~(\ref{eq:TLV}), $|\mu_1| > 1.9 |\mu_3|$ is necessary for $T < 0$. Indeed, as can be seen in Fig.~\ref{fig:T}, if $-2 \alpha T \sim 10^{-3}$ so as to implement the cancellation between tree and loop level contributions, corrections to the light neutrino masses ranging from $\sim 100$~keV to $\sim 100$~MeV, far exceeding present constraints, would be obtained. Thus, we conclude that, while the qualitatively important cancellations described in Ref.~\cite{Akhmedov:2013hec} can in principle take place and affect the constraints on the heavy-active neutrino mixing for $Y_\alpha \sim 1$ and $\Lambda \sim 10$~TeV, in practice large violations of the protecting $B-L$ symmetry would be required, leading to too large radiative corrections to light neutrino masses.    

\section{Conclusions}
\label{sec:sum}

In this work we have analyzed in detail the importance of loop corrections when deriving constraints on the mixing between the SM flavour eigenstates and the new heavy neutrinos introduced in Seesaw mechanisms. Although naively the expectation is that radiative corrections involving these new states would be irrelevant given their weaker-than-weak interactions due to their singlet nature and, a priori, suppressed mixings with the SM neutrinos, Seesaw models allow Yukawa couplings to be sizable, even order one. Thus, loop corrections involving Yukawa vertexes, when the loops involve the heavy neutrinos and the Higgs or the $W$ and $Z$ Goldstones, can indeed be sizable as shown in Ref.~\cite{Akhmedov:2013hec}. In that work, it was shown that, for the low scale Seesaw mechanisms characterized by large Yukawas and low (electroweak) Seesaw scale, the contribution of the new degrees of freedom to the oblique parameters could indeed become as important as the tree level effects in some regions of the parameter space. Moreover, it was observed that several observables shared a common dependence between the $T$ parameter and the tree level contribution, stemming from the modification by these effects of the muon decay through which $G_F$ is determined and subsequently used as input for other observables. Thus, a partial cancellation between these tree and loop level contributions can significantly relax the bounds derived from these observables. Indeed, in Ref.~\cite{Akhmedov:2013hec} a good fit with sizable mixing was obtained in which the most stringent limits were avoided through this partial cancellation while standing tensions between the SM and some observables like the invisible width of the $Z$ were alleviated. 

We have extended the analysis performed in Ref.~\cite{Akhmedov:2013hec} to include also vertex corrections and not only oblique parameters, 
since the sizable contributions from the heavy Yukawas do not vanish when taking the light neutrinos and charged lepton masses to zero. We find that, 
all in all, the oblique parameters do tend to dominate over the other loop corrections and their contribution could be sizable in some part of the parameter space. However, our MCMC scan shows no preference for any sizable loop corrections and the partial cancellation found in~\cite{Akhmedov:2013hec} is not reproduced. We then studied in detail the values of the $T$ parameter preferred by data through our MCMC scan and saw that they were not only negligible, but always positive in our results. Indeed, for the cancellation between tree level contributions and the $T$ parameter to take place, the latter must have negative values. We thus studied the necessary conditions for sizable negative values of the $T$ parameter and realized that, not only sizable Yukawas and relatively low Seesaw scales are required, but also large violations of $B-L$. We then identified the only parameters in the mass matrix with three extra heavy neutrinos that could provide the necessary $B-L$ violation required for $T$ to be negative and competitive with tree level contributions, while keeping neutrino masses within their current bounds despite the large Yukawas, low Seesaw scale and loss of protecting $B-L$ symmetry. Finally, we studied how these parameters would contribute to neutrino masses at loop level and found that, for the size of $T$ required for the cancellation to take place, light neutrino masses would range from 10~keV to 100~MeV, effectively ruling out this possibility.

We conclude that loop level corrections are only relevant in a small fraction of the Seesaw parameter space characterized by large Yukawa couplings and low Seesaw scale and that these corrections tend to strengthen the tree level contributions unless large deviations from $B-L$ are present. If $B-L$ is approximately conserved, data thus prefer regions of the parameter space where these loops are irrelevant. On the other hand, if $B-L$ is strongly violated, the cancellation discussed in Ref.~\cite{Akhmedov:2013hec} can indeed provide a good fit to data with a very relevant role of the loop contributions. However, these large violations of $B-L$ at loop level also lead to too large contributions to the light neutrino masses and hence this possibility is ruled out. We therefore conclude that loop corrections can safely be neglected in analyses of the heavy neutrino mixings in Seesaw mechanisms. 

Finally we have also obtained relevant constraints on this mixing when $B-L$ is an approximate symmetry, so as to recover the correct neutrino masses and mixings observed in neutrino oscillation searches. We find a mild ($\sim$ 90\% CL) preference for non-zero mixing with the $e$ flavours with a best fit at $\theta_e = 0.034^{+0.009}_{-0.014}$ or $\theta_e = 0.035^{+0.009}_{-0.014}$ for normal and inverted mass hierarchy respectively. In the case of normal hierarchy, this preference also induces non-zero mixing with the $\tau$ flavour $\theta_\tau = 0.018^{+0.019}_{-0.013}$ so as to recover the correct pattern of neutrino masses and mixings. On the other hand, small $\theta_\mu$ is preferred so as to keep $\mu \to e \gamma$ at acceptable levels in presence of non-zero $\theta_e$. At the $2 \sigma$ level the following upper bounds are found: $\theta_e < 0.051$, $\theta_\mu < 0.037$ and $\theta_\tau < 0.049$.

\begin{acknowledgments}
We are happy to acknowledge very illuminating discussions with Mattias Blennow, Belen Gavela, Stefano Rigolin and Alfredo Urbano.
We also acknowledge financial support by the European Union through the ITN INVISIBLES (PITN-GA-2011-289442). EFM and JHG also acknowledge 
support from the EU through the FP7 Marie Curie Actions CIG NeuProbes (PCIG11-GA-2012-321582) and the Spanish MINECO through the 
``Ram\'on y Cajal'' programme (RYC2011-07710), the project FPA2009-09017 and through the Centro de excelencia Severo Ochoa Program 
under grant SEV-2012-0249. This work was finalized during the stay of EFM at the Aspen Center for Physics, which is supported by 
the National Science Foundation grant PHY-1066293. This stay was also supported by a grant from the Simons Foundation. ML thanks the IFT UAM/CSIC for the kind hospitality during the early stages of this work. JLP also acknowledges support from the INFN 
program on Theoretical Astroparticle Physics (TASP) and the grant 2012CPPYP7 (Theoretical Astroparticle Physics) under the program 
PRIN 2012 funded by the Italian Ministry of Education, University and Research (MIUR).

\end{acknowledgments}

\section*{Appendix}

In this Appendix we list the self-energies, counterterms and diagrams that enter in the renormalization of the observables studied in Section \ref{sec:obs}. 

\subsection*{Lepton-flavour-dependent counterterms: $\delta^{\text{CT } W}_\alpha$ and $\delta^{\text{CT } Z}$ }

The unrenormalized charged lepton fields $l^0_{L\alpha}$ can be written in terms of the renormalized $\hat{l}_{L\alpha}$ ones as
\begin{equation}
l^0_{L\alpha}=\left(\delta_{\alpha\beta}+\frac{1}{2}\delta Z^{\text{l}}_{\alpha\beta}\right)\hat{l}_{L\beta}.
\end{equation}

The most general expression for the $l_{\beta}\rightarrow l_{\alpha}$ transition amplitude between fermionic Dirac states can 
be written as follows:

\begin{equation}
\Sigma_{\alpha \beta}^{\text{lep}}\left(\slashed{p}\right)=\slashed{p}P_{L}\Sigma_{\alpha \beta}^{L}\left(p^2\right)+
\slashed{p}P_{R}\Sigma_{\alpha \beta}^{R}\left(p^2\right)+P_{L}\Sigma_{\alpha \beta}^{D}\left(p^2\right)+P_{R}
\Sigma_{\alpha \beta}^{D*}\left(p^2\right)\,,
\end{equation}
where $\Sigma^{L}=\Sigma^{L\dagger}$ and $\Sigma^{R}=\Sigma^{R\dagger}$. In the physical observables 
only the Hermitian part of $\delta Z^{\text{l}}$ appears and it is given by
\begin{equation}
\begin{split}
\delta Z^{\text{lep}}_{\alpha \beta}\equiv &\frac{1}{2}\left(\delta Z^{\text{l}}_{\alpha \beta}+\delta Z_{\beta \alpha}^{\text{l}*}\right) \\
=& -\Sigma_{\alpha \beta}^{L}\left(m_{\beta}^{2}\right)-m_{\beta}\Big[ m_{\beta}\Big(\Sigma_{\alpha \beta}^{L\prime}
\left(m_{\beta}^{2}\right)+\Sigma_{\alpha \beta}^{R\prime}\left(m_{\beta}^{2}\right)\Big)+
\Big(\Sigma_{\alpha \beta}^{D\prime}\left(m_{\beta}^{2}\right)+
\Sigma_{\alpha \beta}^{D*\prime}\left(m_{\beta}^{2}\right)\Big) \Big] \, ,
\end{split}
\end{equation}
with $\Sigma^\prime\left(p^{2}\right)\equiv\text{d}\Sigma\left(p^{2}\right)/\text{d}p^{2}$. Therefore, the heavy neutrino  
contribution to $\delta Z^{\text{lep}}$ can be obtained simply computing

\begin{picture}(550,60)
\Line[arrow,arrowpos=0.5,arrowlength=5,arrowwidth=2,arrowinset=0.2](0,0)(60,0)
\Arc[dash,dashsize=8](90,0)(30,0,-180)
\Line(60,0)(120,0)
\Line[arrow,arrowpos=0.5,arrowlength=5,arrowwidth=2,arrowinset=0.2](120,0)(180,0)
\put(30,-14){$l_\alpha^{\pm}$}
\put(150,-14){$l_\beta^{\pm}$}
\put(88,-14){$N_k$}
\put(88,38){$\phi^{\pm}$}
\put(200,-2){$= i\Sigma_{\alpha \beta}^{\text{lep}}(\slashed{p}) \Rightarrow$ }
\end{picture}

\vspace{0.2cm} 

\begin{equation}
\begin{split}
\Sigma_{\alpha \beta}^{\text{lep}}(\slashed{p})=&-\frac{\alpha}{8\pi s_\mathrm{W}^2 M_W^2}\sum_{k=4}^6 \bigg\lbrace M_{k}^2 U_{\beta k} U^*_{\alpha k} \Big[\left( P_L m_{\beta}+P_R m_{\alpha}\right) B_0(p^2,M_{k}^2,M_W^2) \\
&+\slashed{p} \left(P_R \frac{m_{\alpha} m_{\beta}}{M_{k}^2}+P_L\right) B_1(p^2,M_{k}^2,M_W^2) \Big]\bigg\rbrace \, ,
\end{split}
\end{equation}
where $B_i$ (and later $B_{ij}$, $C_{ij}$, $D_i$ and $D_{ij}$) are the Passarino-Veltman integrals~\cite{Passarino:1978jh} using the notation from Ref.~\cite{Ellis:2011cr}.

Similarly, the unrenormalized neutrino fields $\nu^0_{Lj}$ can also be written in terms of the renormalized ones 
$\hat{\nu}_{Lj}$ as
\begin{equation}
\nu^0_{Li}=\left(\delta_{ij}+\frac{1}{2}\delta Z^{\nu}_{ij}\right)\hat{\nu}_{Lj}.
\end{equation}

The transition amplitude between two Majorana states reads

\begin{equation}
\Sigma_{ij}^\text{neu}\left(\slashed{p}\right)=\slashed{p}P_{L}\Sigma_{ij}^{L}\left(p^2\right)+\slashed{p}P_{R}
\Sigma_{ij}^{L*}\left(p^2\right)+P_{L}\Sigma_{ij}^{M}\left(p^2\right)+P_{R}\Sigma_{ij}^{M*}\left(p^2\right) \, ,
\end{equation}
where $\Sigma^{L}=\Sigma^{R *}$ and $\Sigma^{M}=\Sigma^{M t}$. In the Majorana case, the Hermitian part of $\delta Z^{\nu}$ can be 
written as
\begin{equation}
\begin{split}
\delta Z^{\text{neu}}_{ij}\equiv&\frac{1}{2}\left(\delta Z^{\nu}_{ij}+\delta Z^{\nu *}_{ji}\right) \\
=&-\Sigma_{ij}^{L}\left(m_{j}^{2}\right)-m_{j}\Big[ m_{j}\Big(\Sigma_{ij}^{L\prime}\left(m_{j}^{2}\right)+\Sigma_{ij}^{L*\prime}\left(m_{j}^{2}\right)\Big)+\Big(\Sigma_{ij}^{M\prime}\left(m_{j}^{2}\right)+\Sigma_{ij}^{M*\prime}\left(m_{j}^{2}\right)\Big) \Big]\, .
\end{split}
\end{equation}
Analogously to the charged lepton case, $\delta Z^{\text{neu}}$ can thus be obtained from the heavy neutrino contribution 
to the neutrino self energy:

\begin{picture}(550,60)
\Line[arrow,arrowpos=0.5,arrowlength=5,arrowwidth=2,arrowinset=0.2](0,0)(60,0)
%\Arc[clock](90,0)(30,0,180)%
\Arc[dash,dashsize=8](90,0)(30,0,-180)
\Line(60,0)(120,0)
\Line[arrow,arrowpos=0.5,arrowlength=5,arrowwidth=2,arrowinset=0.2](120,0)(180,0)
\put(30,-14){$\nu_i$}
\put(150,-14){$\nu_j$}
\put(88,-14){$N_k$}
%\put(88,-23){$n_k$}%
\put(80,38){$\phi^{0},\, H$}
\put(200,-2){$= i\Sigma_{ij}^{\text{neu}}(\slashed{p}) \Rightarrow$ }
\end{picture}

\vspace{0.2cm} 

\begin{equation}
\begin{split}
\Sigma_{ij}^{\text{neu}}(\slashed{p})=&-\frac{\alpha}{16\pi s_\mathrm{W}^2 M_W^2} \sum_{k=4}^6 \bigg\lbrace \slashed{p} P_L \left(M_j C_{jk}^*+M_k C_{jk}\right) \left(M_iC_{ki}^*+M_k C_{ki}\right)  \\
& \times \Big[B_1(p^2,M_k^2,M_Z^2)+B_1(p^2,M_k^2,M_h^2)\Big]  \\
& +\slashed{p} P_R  \left(M_j C_{jk}+M_k C^*_{jk} \right) \left(M_iC_{ki}+ M_kC^*_{ki}\right)  \\
&  \times \Big[B_1(p^2,M_k^2,M_Z^2)+B_1(p^2,M_k^2,M_h^2)\Big]\\
& +P_L M_k \left(M_jC_{jk}+M_kC^*_{jk}\right)\left(M_kC_{ki}+M_iC^*_{ki}\right)  \\
& \times \Big[B_0(p^2,M_k^2,M_Z^2)-B_0(p^2,M_k^2,M_h^2)\Big]\\
& +P_R M_k\left(M_j C^*_{jk}+M_kC_{jk}\right)\left(M_kC^*_{ki} +M_iC_{ki}\right)  \\
& \times \Big[B_0(p^2,M_k^2,M_Z^2)-B_0(p^2,M_k^2,M_h^2)\Big] \bigg\rbrace \, .
\end{split}
\end{equation}

Finally, the lepton-flavour-dependent combinations that will correct and cancel the divergences of 1-loop corrections to the vertex $W\nu l_\alpha$ and $Z\nu\nu$ are respectively:

\begin{eqnarray}
\label{eq:propscorr}
\delta^{\text{CT } W}_\alpha &=& \displaystyle\sum_{i=1}^3{\frac{U_{\alpha i}}{2}\left( \displaystyle\sum_{\beta=1}^3{\delta Z^{\text{lep}}_{\beta \alpha}U^*_{\beta i}}+\displaystyle\sum_{j=1}^6{U^*_{\alpha j}\delta Z^{\text{neu}}_{i j}}\right)} \\
\delta^{\text{CT } Z} &=& \displaystyle\sum_{k=1}^6 {\left( \delta Z^{\text{neu}}_{i k}C_{kj}+\delta Z^{\text{neu}}_{k j}C_{i k} \right)}
\label{eq:propscorrZ}
\end{eqnarray}

\subsection*{Vertex interferences: $\mathcal{V}_\alpha^W$ and $\mathcal{V}_{ij}^Z$}

\begin{picture}(550,40)
\Photon(0,0)(60,0){5}{3}
\Line(120,-30)(120,30)
\Line[dash,dashsize=8](60,0)(120,30)
\Line[dash,dashsize=8](60,0)(120,-30)
\Line[arrow,arrowpos=0.5,arrowlength=5,arrowwidth=2,arrowinset=0.2](120,30)(180,30)
\Line[arrow,arrowpos=0.5,arrowlength=5,arrowwidth=2,arrowinset=0.2](180,-30)(120,-30)
\put(30,14){$W_{\mu}^{\pm}$}
\put(85,-35){$h,\, \phi^0$}
\put(90,30){$\phi^\pm$}
\put(126,-2){$N_{k}$}
\put(190,-30){$\overline{\nu}_{i}$}
\put(190,30){$l_{\alpha}^\pm$}
\put(200,-2){$= i T^{V}_{W\nu_il_\alpha} \Rightarrow$ }
\end{picture}

\vspace{1cm} 

\begin{equation}
\begin{split}
\mathcal{V}_\alpha^W\equiv&\displaystyle \frac{\sum_{i=1}^3 T_{0}^*T^{V}_{W\nu_il_\alpha}}{\sum_{i=1}^3 |T_0|^2} \\
=&\frac{\alpha}{8\pi s_\mathrm{W}^2 M_W^2}\displaystyle\sum_{i=1}^3\displaystyle\sum_{k=4}^6{M_{k}^2U_{\alpha i}U^*_{\alpha k}C^*_{ki}\left[ C_{00}(0,0,M_h^2,M_{k}^2,M_W^2)+C_{00}(0,0,M_Z^2,M_{k}^2,M_W^2)\right]} ,
\end{split}
\label{eq:Wvertex}
\end{equation}
up to higher order corrections and where $T_0$ is the corresponding tree level amplitude.

%%%%%%%%%%%%%%%

\vspace{1cm}
\begin{picture}(550,40)
\Photon(0,0)(60,0){5}{3}
\Line(120,-30)(120,30)
\Line[dash,dashsize=8](60,0)(120,30)
\Line[dash,dashsize=8](60,0)(120,-30)
\Line[arrow,arrowpos=0.5,arrowlength=5,arrowwidth=2,arrowinset=0.2](120,30)(180,30)
\Line[arrow,arrowpos=0.5,arrowlength=5,arrowwidth=2,arrowinset=0.2](180,-30)(120,-30)
\put(180,0){$+$}
\Photon(200,0)(260,0){5}{3}
\Line[dash,dashsize=8](320,-30)(320,30)
\Line(260,0)(320,30)
\Line(260,0)(320,-30)
\Line[arrow,arrowpos=0.5,arrowlength=5,arrowwidth=2,arrowinset=0.2](320,30)(380,30)
\Line[arrow,arrowpos=0.5,arrowlength=5,arrowwidth=2,arrowinset=0.2](380,-30)(320,-30)
\put(30,14){$Z_{\mu}$}
\put(90,-35){$h$}
\put(90,30){$\phi^0$}
\put(126,-2){$N_{k}$}
\put(190,-30){$\overline{\nu}_{i}$}
\put(190,30){$\nu_j$}
\put(230,14){$Z_{\mu}$}
\put(290,-35){$N_r$}
\put(290,30){$N_k$}
\put(330,-2){$h,\, \phi^0$}
\put(390,-30){$\overline{\nu}_{i}$}
\put(390,30){$\nu_j$}
\put(400,-2){$= i T^{V}_{Z\nu_i\nu_j} \Rightarrow$ }
\end{picture}
\vspace{1cm} 

\begin{equation}
\begin{split}
\mathcal{V}_{ij}^Z \equiv & \frac{T_{0}^*T^{V}_{Z\nu_i\nu_j}}{|T_0|^2}\\
= & \frac{\alpha}{16\pi s_\mathrm{W}^2 M_W^2}\Bigg[\displaystyle\sum_{k,r=4}^6 \Bigg\lbrace -2 C_{kj}C_{ir}M_k M_r \bigg(C_{rk}M_kM_r\Big[C_0(0,M_Z^2,M_h^2,M_k^2,M_r^2)\\
&+C_0(0,M_Z^2,M_Z^2,M_k^2,M_r^2)\Big]+C_{kr}\Big[M_Z^2\big(C_{22}(0,M_Z^2,M_h^2,M_k^2,M_r^2)\\
&+C_{22}(0,M_Z^2,M_Z^2,M_k^2,M_r^2)-C_{21}(0,M_Z^2,M_h^2,M_k^2,M_r^2)-C_{21}(0,M_Z^2,M_Z^2,M_k^2,M_r^2)\big)\\ 
&+2\big(C_{00}(0,M_Z^2,M_Z^2,M_k^2,M_r^2)+C_{00}(0,M_Z^2,M_h^2,M_k^2,M_r^2)\big)\Big]\bigg)\Bigg\rbrace\\
&+\displaystyle\sum_{k=4}^6\Big[4 C_{kj}C_{ik}M_k^2\big(C_{00}(0,M_Z^2,M_k^2,M_h^2,M_Z^2)+C_{00}(0,M_Z^2,M_k^2,M_Z^2,M_h^2)\big)\Big]\Bigg]\, ,\end{split}
\label{eq:Zvertex}
\end{equation}
up to higher order corrections and where $T_0$ is the corresponding tree level amplitude.

\subsection*{Box contribution to $\mu$ decay: $\mathcal{B}_{\alpha \beta}$}

\begin{picture}(550,50)
\Line[arrow,arrowpos=0.5,arrowlength=5,arrowwidth=2,arrowinset=0.2](0,50)(40,30)
\Line(40,30)(40,-30)
\Line[arrow,arrowpos=0.5,arrowlength=5,arrowwidth=2,arrowinset=0.2](40,-30)(80,-50)
\Line[dash,dashsize=8](40,30)(100,30)
\Line[dash,dashsize=8](40,-30)(100,-30)
\Line(100,30)(100,-30)
\Line[arrow,arrowpos=0.5,arrowlength=5,arrowwidth=2,arrowinset=0.2](100,30)(140,50)
\Line[arrow,arrowpos=0.5,arrowlength=5,arrowwidth=2,arrowinset=0.2](140,-50)(100,-30)
\put(160,-2){$= i T^{B}_{\alpha \beta} \Rightarrow$ }
\put(15,25){$l^\pm_\alpha$}
\put(120,25){$l^\pm_\beta$}
\put(85,-50){${\nu_j}$}
\put(145,-50){${\overline{\nu}_i}$}
\put(65,34){$\phi^\pm$}
\put(55,-26){$\phi^0\,,h$}
\put(102,-2){$N_k$}
\put(25,-2){$N_r$}
\end{picture}

\vspace{1.5cm} 

\begin{equation}
\begin{split}
\mathcal{B}_{\alpha \beta} \equiv & \frac{\sum_{i,j=1}^3T_{0}^*T^{B}_{\alpha \beta}}{\sum_{i,j=1}^3|T_{0}|^2} \\
= & \frac{1}{5}\frac{g^2}{(16\pi)^2 M_W^2} \displaystyle\sum_{i,j=1}^3\displaystyle\sum_{k,r=4}^6{C_{ik}C_{jr}U_{\beta k}U^*_{\beta i}U^*_{\alpha r}U_{\alpha j}M^2_r M^2_k} \bigg\lbrace 20\Big[D_{00}(M_h^2)+D_{00}(M_Z^2)\Big] \\
&+m_\alpha^2\Big[3\big(D_{12}(M_h^2)+D_{12}(M_Z^2)\big)+2\big(D_{13}(M_h^2)+D_{13}(M_Z^2)\big)\\
&+3\big(D_{2}(M_h^2)+D_{2}(M_Z^2)\big) +2\big(D_{3}(M_h^2)+D_{3}(M_Z^2)\big)\Big] \bigg\rbrace\, ,
\end{split}
\label{eq:box}
\end{equation}
up to higher order corrections and where $T_0$ is the corresponding tree level amplitude and using the simplified notation $D_{ij}(M^2) \to D_{ij}(0,0,0,M_r^2,M^2,M_k^2,M_W^2)$. Apart from the explicit sum over final state neutrinos in Eq.~(\ref{eq:box}), the integral over the phase space is to be understood in both the numerator and denominator.

\subsection*{Flavour-universal corrections to the gauge boson propagators: $\delta_W^\text{univ N}$ and $\delta_Z^\text{univ N}$}

We label $\Sigma_{WW}$ and $\Sigma_{ZZ}$ the terms proportional to $g^{\mu\nu}$ in the $W$ and $Z$ self-energies respectively. Notice that the SM contribution has been subtracted from the total self-energy, as we are interested in the contribution stemming from the new extra neutrinos.

\begin{picture}(550,60)
\Photon(0,0)(60,0){5}{3}
\Arc(90,0)(30,0,360)
\Photon(120,0)(180,0){5}{3}
\put(30,14){$W^{\pm}$}
\put(150,14){$W^{\pm}$}
\put(84,-44){$N_i$}
\put(84,38){$l_{\alpha}^{\pm}$}
\put(200,-2){$= i\Sigma_{WW}^\text{tot}(p^2) \Rightarrow$}\\
\end{picture}

\vspace{1cm} 

\begin{equation}
\begin{split} 
\Sigma_{WW}^N(p^2)\equiv&\Sigma_{WW}^\text{tot}(p^2)-\Sigma_{WW}^\text{SM}(p^2) \\
=&-\frac{\alpha}{4\pi s_\mathrm{W}^2}\displaystyle\sum_{\alpha=e,\mu,\tau}\Bigg\lbrace \displaystyle\sum_{i=1}^6 \vert U_{\alpha i}\vert^2 \bigg[ 2 B_{00}(p^2,M_i^2,m_\alpha^2) +p^2\Big( B_1(p^2,M_i^2,m_\alpha^2)  \\
 &+ B_{11}(p^2,M_i^2,m_\alpha^2)\Big) \bigg] - 2 B_{00}(p^2,0,m_\alpha^2)-p^2\Big(B_1(p^2,0,m_\alpha^2)+B_{11}(p^2,0,m_\alpha^2)\Big)\Bigg\rbrace 
\end{split}
\label{eq:sigmaW}
\end{equation}

\begin{picture}(550,60)
\Photon(0,0)(60,0){5}{3}
\Arc(90,0)(30,0,360)
\Photon(120,0)(180,0){5}{3}
\put(30,14){$Z$}
\put(150,14){$Z$}
\put(84,38){$N_i$}
\put(84,-44){$N_{j}$}
\put(200,-2){$= i\Sigma_{ZZ}^\text{tot}(p^2) \Rightarrow$ }
\end{picture}

\vspace{1cm} 

\begin{equation}
\begin{split}
\Sigma_{ZZ}^N(p^2)\equiv&\Sigma_{ZZ}^\text{tot}(p^2)-\Sigma_{ZZ}^\text{SM}(p^2)\\
=&-\frac{\alpha}{8\pi s_\mathrm{W}^2 c_\mathrm{W}^2}\Bigg\lbrace\displaystyle\sum_{\alpha, \beta}\displaystyle\sum_{i,j=1}^6 \bigg[ U_{\alpha i}U_{\alpha j}^*U_{	\beta j}U_{\beta i}^*M_iM_jB_0(p^2,M_i^2,M_j^2) + U_{\alpha j}U_{\alpha i}^*U_{\beta i}U_{\beta j}^*\\
&\times\Big( 2 B_{00}(p^2,M_i^2,M_j^2)+p^2\big(B_1(p^2,M_i^2,M_j^2)+B_{11}(p^2,M_i^2,M_j^2)\big)\Big) \bigg] \\
&- 3\Big[ 2 B_{00}(p^2,0,0)+p^2\big(B_1(p^2,0,0)+B_{11}(p^2,0,0)\big)\Big]\Bigg\rbrace
\end{split}
\label{eq:sigmaZ}
\end{equation}
Notice that both in Eq.~(\ref{eq:sigmaW}) and in Eq.~(\ref{eq:sigmaZ}) the sums run over all neutrino mass eigenstates (heavy and light) so here $M_i$ can represent both the heavy or the light neutrino masses.

The oblique universal corrections to the electroweak observables can be written as a combination of the three following 
independent parameters~\cite{Peskin:1990zt,Peskin:1991sw}:

\begin{eqnarray}
\label{S}
\alpha S&=&\frac{4s_\mathrm{W}^{2}c_\mathrm{W}^{2}}{M_{Z}^{2}}\Bigl[\hat{\Sigma}_{ZZ}^N(0)
+\hat{\Sigma}_{\gamma \gamma}^N(M_{Z}^{2})
-\frac{c_\mathrm{W}^{2}-s_\mathrm{W}^{2}}{c_\mathrm{W}s_\mathrm{W}}\hat{\Sigma}_{Z\gamma}^N(M_{Z}^{2})\Bigr]\,,\\
\label{T}
\alpha T&=&\frac{\hat{\Sigma}_{ZZ}^N(0)}{M_{Z}^{2}}-\frac{\hat{\Sigma}_{WW}^N(0)}{M_{W}^{2}}\,,\\
\label{U}
\alpha U&=&4s_\mathrm{W}^{2}c_\mathrm{W}^{2}\biggl[\frac{1}{c_\mathrm{W}^{2}}\frac{\hat{\Sigma}_{WW}^N(0)}{M_{W}^{2}}-
\frac{\hat{\Sigma}_{ZZ}^N(0)}{M_{Z}^{2}}
+\frac{s_\mathrm{W}^{2}}{c_\mathrm{W}^{2}}\frac{\hat{\Sigma}_{\gamma \gamma}^N(M_{Z}^{2})}{M_{Z}^{2}}-
\frac{2s_\mathrm{W}}{c_\mathrm{W}}\frac{\hat{\Sigma}_{Z\gamma}^N(M_{Z}^{2})}{M_{Z}^{2}}\biggr]\,.
\end{eqnarray}
and the renormalized self energies are given by:

\begin{eqnarray}
\hat{\Sigma}_{WW}^N\left(p^2\right)&=&\Sigma_{WW}^N\left(p^2\right)-\Sigma_{WW}^N\left(M_W^2\right)+(p^2-M_W^2)\left[
\frac{c_\mathrm{W}^2}{s_\mathrm{W}^2}\mathcal{R} -\Sigma_{\gamma\gamma}^{N\prime}(0)\right],\nonumber\\
\hat{\Sigma}_{ZZ}^N\left(p^2\right)&=&\Sigma_{ZZ}^N\left(p^2\right)-\Sigma_{ZZ}^N\left(M_Z^2\right)+(p^2-M_Z^2)\left[
\left(\frac{c_\mathrm{W}^2}{s_\mathrm{W}^2}-1\right)\mathcal{R}-\Sigma_{\gamma\gamma}^{N\prime}(0)\right],\nonumber\\
\hat{\Sigma}_{Z\gamma}^N\left(p^2\right)&=& \Sigma_{Z\gamma}^N\left(p^2\right)-\Sigma_{Z\gamma}^N\left(0\right)-p^2\frac{c_\mathrm{W}}{s_\mathrm{W}}\mathcal{R},\nonumber\\
\hat{\Sigma}_{\gamma\gamma}^N\left(p^2\right)&=& \Sigma_{\gamma\gamma}^N\left(p^2\right)-p^2\Sigma_{\gamma\gamma}^{N\prime}\left(0\right),
\end{eqnarray}
with 
\be
\mathcal{R}=\frac{\Sigma_{ZZ}^N\left(M_Z^2\right)}{M_Z^2}-\frac{\Sigma_{WW}^N\left(M_W^2\right)}{M_W^2}-\frac{2s_\mathrm{W}}{c_\mathrm{W}}
\frac{\Sigma_{Z\gamma}^N\left(0\right)}{M_Z^2}
\ee
Notice that, in the on-shell renormalization scheme $\hat{\Sigma}_{WW}^N\left(M_W^2\right)=\hat{\Sigma}_{ZZ}^N\left(M_Z^2\right)=\hat{\Sigma}_{Z\gamma}^N\left(0\right)
=\hat{\Sigma}_{\gamma\gamma}^N\left(0\right)=0$. Moreover, there is no contribution to the propagator of the photon 
from the extra heavy neutrinos and therefore $\Sigma_{\gamma\gamma}^N$ and $\hat{\Sigma}_{\gamma\gamma}^N$ can be set to zero in the previous equations. In addition,
there is no correction to $\Sigma_{Z\gamma}$ either, so that $\Sigma_{Z\gamma}^N$ can be set to zero too. The universal oblique counterterms presented in Sec.~\ref{sec:obs} can thus be written as:

\begin{eqnarray}
\delta_W^\text{univ N}&=&\frac{\Sigma_{WW}^N\left(0 \right)-\Sigma_{WW}^N\left(M_W^2\right)}{M_W^2}-\frac{c_\mathrm{W}^2}{s_\mathrm{W}^2}\mathcal{R}=
\frac{\hat{\Sigma}_{WW}^N\left(0\right)}{M_W^2}\nonumber\\
\label{dW}
&=& \frac{1}{2s_\mathrm{W}^2}\alpha S-\frac{ c_\mathrm{W}^2}{s_\mathrm{W}^2}\alpha T-\frac{\cos 2\theta_W}{4s_\mathrm{W}^4}\alpha U\nonumber\\
\delta_Z^\text{univ N}&=&\frac{\Sigma_{ZZ}^N\left(0 \right)-\Sigma_{ZZ}^N\left(M_Z^2\right)}{M_Z^2} + 
\frac{1}{2}\left(1-\frac{c_\mathrm{W}^2}{s_\mathrm{W}^2}\right)\mathcal{R}=\frac{\hat{\Sigma}_{ZZ}^N\left(0\right)}{M_Z^2}\nonumber\\
\label{dZ}
&=&\frac{1}{2s_\mathrm{W}^2}\alpha S+\left(1-\frac{c_\mathrm{W}^2}{s_\mathrm{W}^2}\right)\alpha T-\frac{\cos 2\theta_W}{4s_\mathrm{W}^4}\alpha U .
\end{eqnarray}

\end{document}